\documentclass[authorversion, nonacm]{acmart}
\AtBeginDocument{%
  }

\setcopyright{none}
\copyrightyear{2025}
\acmYear{2025}
\acmDOI{XXXXXXX.XXXXXXX}
\acmConference[FAccT '25]{ACM Conference on Fairness, Accountability, and Transparency}{June 23--26,
  2025}{Athens, Greece}
\acmISBN{978-1-4503-XXXX-X/18/06}

\usepackage{multirow}
\usepackage{booktabs} 
\usepackage{soul}
\usepackage{geometry}
\usepackage{array}
\usepackage{longtable}
\usepackage{graphicx}
\usepackage{diagbox}
\begin{document}

\title{Envisioning Stakeholder-Action Pairs to Mitigate Negative Impacts of AI: A Participatory Approach to Inform Policy Making}

\author{Julia Barnett}
\email{JuliaBarnett@u.northwestern.edu}
\orcid{0000-0002-3476-1110}
\affiliation{%
  \institution{Northwestern University}
  \city{Evanston}
  \state{IL}
  \country{USA}}

\author{Kimon Kieslich}
\email{k.kieslich@uva.nl}
\orcid{0000-0002-6305-2997}
\affiliation{%
  \institution{University of Amsterdam}
  \city{Amsterdam}
  \country{The Netherlands}}
\email{k.kieslich@uva.nl}
\orcid{0000-0002-6305-2997}

\author{Natali Helberger}
\email{n.helberger@uva.nl}
\orcid{0000-0003-1652-0580}
\affiliation{%
  \institution{University of Amsterdam}
  \city{Amsterdam}
  \country{The Netherlands}}

\author{Nicholas Diakopoulos}
\email{nad@northwestern.edu}
\orcid{0000-0001-5005-6123}
\affiliation{%
  \institution{Northwestern University}
  \city{Evanston, IL}
  \country{USA}}

\setcopyright{none}

\renewcommand{\shortauthors}{Barnett et al.}
\renewcommand{\shorttitle}{Envisioning Action-Stakeholder Pairs to Mitigate Negative Impacts of AI}
\begin{abstract}
  The potential for negative impacts of AI has rapidly become more pervasive around the world, and this has intensified a need for responsible AI governance. While many regulatory bodies endorse risk-based approaches and a multitude of risk mitigation practices are proposed by companies and academic scholars, these approaches are commonly expert-centered and thus lack the inclusion of a significant group of stakeholders. Ensuring that AI policies align with democratic expectations requires methods that prioritize the voices and needs of those impacted.  
  In this work we develop a participative and forward-looking approach to inform policy-makers and academics that grounds the needs of lay stakeholders at the forefront and enriches the development of risk mitigation strategies. Our approach (1) maps potential mitigation and prevention strategies of negative AI impacts that assign responsibility to various stakeholders, (2) explores the importance and prioritization thereof in the eyes of laypeople, and (3) presents these insights in policy fact sheets, i.e., a digestible format for informing policy processes. We emphasize that this approach is not targeted towards replacing policy-makers; rather our aim is to present an informative method that enriches 
  mitigation strategies and enables a more participatory approach to policy development.
\end{abstract}






\maketitle

\section{Introduction}

The rapid development of AI systems that have become commonplace in the daily lives of millions of people around the world has spurred a growing need for responsible AI governance. More broadly, this has resulted in a proliferation of frameworks, guides, and principles all detailing best practices for using, designing, and deploying these AI systems. In recent years, there have been various regulatory endeavors to temper and prevent harms of AI systems at federal or even multinational levels with measures like Biden's Executive Order on Artificial Intelligence in the United States \cite{biden_exec_order}, the EU AI Act \cite{eu_ai_act_52}, or the Chinese Generative AI Regulation \cite{chinese_gen_ai_regulation}. Most of these approaches are risk-based, i.e., they aim to identify negative impacts that can materialize through AI systems and then propose mitigation measures \cite{van_der_heijden_risk_2021}. One central aspect for the effectiveness of these approaches is to critically interrogate who is conducting the risk analysis and subsequent risk mitigation measures. Currently, the efforts in AI governance, whether official regulatory measures, academic works to create guidelines, or sets of standards established by technology companies, all tend to rely on a small set of expert knowledge \cite{stahl_systematic_2023}. 

Thus, a problem emerges---how to integrate and incorporate input from broader sets of stakeholders in order to develop policy consistent with democratic expectations. There is value in lay stakeholder (a subsample of the public with no formal or professional link to the technology at hand) knowledge, since they can report on lived experiences, describe how negative impacts materialize in a diversity of real-world settings that they inhabit, and formulate expectations based on their underlying values and norms for how these impacts might be mitigated \cite{moss_assembling_2021, diaz_whose_2019, kieslich_anticipating_2024}. Consequently, incorporating expertise and experience from laypeople offers regulators and policy makers a greater sense of what people actually expect. Such input is useful for policy makers so they have guidance and an empirical grounding to incorporate more diverse stakeholder opinions into their development process. Following this participatory process can also build trust and lead to stronger alignment of decision maker goals with the expectations of laypeople.

One methodological challenge for integrating knowledge from lay stakeholders is contextualizing the often complex set of AI impacts so that they can clearly understand the issues and effectively weigh in. An approach to this which we have explored in prior work \cite{barnett2024simulating, kieslich_anticipating_2024} is to narrativize these complex impacts through the use of written scenarios---short stories specifically designed to illustrate the negative impacts of AI systems. Here, we leverage this approach for presenting a broader set of laypeople with narrativized negative AI impacts to help identify mitigation measures and responsibilities of actors to implement those measures. Thus, our approach helps in streamlining the complex risk mitigation process in tapping into the perceptions of a broader set of stakeholders and gathering their insights for brainstorming and prioritization of possible actions and responsibility allocations. 

In this paper we present an approach to systematically collect potential mitigation strategies for current and future negative impacts of an emerging technology, here focusing on the impacts of generative AI in the media environment
\cite{kieslich_anticipating_2024}.
This approach not only serves to help identify potential actions to mitigate impact, but also solicits responsibility allocations reflecting the priorities of lay respondents. Concretely, we present participants in a survey (Survey 1) with various scenarios for generative AI's future impact on the media environment and use these stimuli as an example to identify mitigation approaches, i.e., \textit{what} should be done \textit{by whom} to address negative impacts. We then engage a different set of participants (Survey 2) to prioritize those mitigation strategies. Finally, we utilize a large language model (LLM) to create policy fact sheets, i.e., short summaries of the highly prioritized action stakeholder pairs. Our aim is to (1) inform policy-makers and academics in the mapping of potential mitigation strategies and responsibilities of various stakeholders in the eyes of laypeople, (2) guide the prioritization of those measures, and (3) present short fact sheets based on lay input that can function as empirically-grounded fodder for policy makers. 



\section{Related Work}

To position our approach within the literature, we first outline the state of AI governance and critically assess the risk-based approaches typically used. We highlight the lack of input from lay stakeholders in risk assessment and mitigation approaches and elaborate how participatory methods help address this gap and can enhance policy formulation.

\subsection{Governance of AI}

Driven by massive investments in AI and the strategic race for power, AI systems are being rapidly (openly) released, with huge implications for individuals, organizations, and society. These impacts can be both positive and negative. As a result, governments around the world have adopted risk-based approaches that introduce various forms of AI policy that seek to balance the protection of societal values while enabling innovation \cite{ezeani_survey_2021, van_der_heijden_risk_2021}. At the academic level (often co-authored by tech company researchers  \cite{weidinger2023sociotechnical, weidinger2022taxonomy}), many scholars have introduced their own impact assessment frameworks that aim to identify potential harms to their systems and outline actions to mitigate them \cite{baldassarre_social_2023, bird_typology_2023, hoffmann_adding_2023, shelby_sociotechnical_2023, slattery_ai_2024, solaiman_evaluating_2023, stahl_systematic_2023, zeng_ai_2024}. In addition, standardization bodies publish safety guidelines on how to address the harms of AI systems \cite{nist_artificial_2024}. In 2024, the EU brought the EU AI Act into force \cite{eu_ai_act_52}, which legally mandates risk assessments for AI systems, and the United States saw similar regulation with the Executive Order 14110: Executive Order on Artificial Intelligence \cite{biden_exec_order}.

However, while acknowledging the variety of different policy and security frameworks, it is necessary to critically question the efficacy of methods of impact identification and subsequent mitigation strategies to actually protect a diverse citizenry. The emphasis here is on the diversity of the citizenry and the impact of technology on them. This is even more pronounced when considering that end-users are often directly involved with AI technologies and can actively shape the direction of technology development and implementation. Thus, a first critical factor in determining the effectiveness of AI governance proposals is to examine who is actively engaged in defining algorithmic impacts. Scholars have argued \cite{van_der_heijden_risk_2021} that the objective nature of current assessments should be questioned, as identifying impacts and developing mitigation strategies are political and contextual decisions \cite{orwat_normative_2024, van_der_heijden_risk_2021}. Often these decisions are in the hands of experts or companies themselves, which can be biased \cite{nanayakkara_unpacking_2021-1, schmitz_global_2024, hartmann_addressing_2024, bonaccorsi_expert_2020}. Even further, it is debatable how effective impact assessments are when they are conducted by companies whose primary goal is to profit from the technology \cite{mittelstadt_principles_2019, stahl_systematic_2023, waldman_privacy_2019}. Thus, while many scholars emphasize the importance of broader inclusion \cite{metcalf_algorithmic_2021, moss_assembling_2021, mesmer_auditing_2023, barnett2022crowdsourcing}, in practice this is critically lacking, especially with respect to ordinary citizens and civil society actors \cite{griffin_what_2024, gillespie_generative_2024, kieslich_using_2024}.  

A second critical factor to consider is the ability of current governance approaches to adapt to new AI applications that may have novel impacts. Currently, the regulatory framework in the EU (EU AI Act; Digital Service Act (DSA)) requires companies to address reasonably foreseeable risks or ``known knowns'' \cite{ebers_truly_2024}. Accordingly, most existing assessment and mitigation frameworks focus on predefined checklists or questionnaires based on known impact criteria \cite{stahl_systematic_2023, orwat_normative_2024}. This focus leaves little room for the detection of as-yet-unknown impacts that, for example, may only emerge through end-user interaction with the technology, although there is promise in using AI technology to envision potential impacts and uses of these tools \cite{herdel2024exploregen}. These novel impacts might also require new policy approaches as the current ones might not be adequately fitted to counter potential harms. But while more qualitative methods that enable the identification of contextualized risks and their mitigation are needed \cite{gellert_risk-based_2020}, the current regulatory landscape is more reactive and relies on established (quantitative) parameters.   

As a result, the vast majority of current assessment and mitigation methods are inadequate because they lack context, require simplification of societal phenomena, and fail to capture the breadth of potential impacts \cite{gellert_risk-based_2020, orwat_normative_2024, pasquale_power_2023, walker_identifying_2024}. For example, simple compliance procedures may make an AI system safer in some forms, but may not be able to respond to changes in the environment. In addition, a focus on quantification might leave out more qualitative types of data, such as the stories or lived experiences of those affected by the technology. These lived experiences in particular can make impacts more tangible and also help to develop citizen-centered governance structures that resonate with the public \cite{diaz_whose_2019, moss_assembling_2021}. Furthermore, most of the current forms of governance are slow and reactive to the technology that has already been implemented. As such, there is a need for methods that accelerate this process and provide a forward-looking view of potential future developments as well as what actions need to be taken by whom to avert negative consequences. 

\subsection{Participatory Approaches for Envisioning Risk Mitigation Actions}

In response to the top-down, expert-driven practice of current regulatory and operational risk-based approaches, scholars argue for a more inclusive and participatory approach to address the challenges posed by AI systems \cite{metcalf_algorithmic_2021, moss_assembling_2021, gellert_risk-based_2020}. This also requires a change in approach to more qualitative forms of assessments in terms of thinking about plausible future developments \cite{gellert_risk-based_2020, pasquale_power_2023}. These anticipatory efforts have proven fruitful in the past: the anticipatory governance literature discusses approaches to identify and mitigate adverse impacts of technology at an early stage \cite{guston_understanding_2013, brey_anticipatory_2012}. The main goal of those studies is to identify negative impacts early in the development and implementation phase of emerging technologies and subsequently propose mitigation strategies before these impacts materialize \cite{fuerth_operationalizing_2011, guston_understanding_2013, selbst_institutional_2021}. By illuminating future pathways and proposing different policy options, anticipatory governance studies empirically provide a deliberative space to navigate the risk mitigation decision-making process \cite{mittelstadt_how_2015, guston_understanding_2013}. 

One such methodological approach to discovering potential impacts of technology is scenario writing or scenario planning \cite{amer_review_2013, borjeson_scenario_2006, andersen_stakeholder_2021, ramirez_strategic_2016}, which has also been identified by standards organizations as a viable method to support future thinking \cite{international_organization_for_standardization_risk_2019}. Scenarios allow decision makers to think through different plausible futures and discuss strategies for mitigating potential harm or even aiming for a desired future. In this method, scientists highlight the importance of different stakeholders (e.g., experts, stakeholder representatives, citizens, etc.) early on in the scenario planning process. It stimulates engagement with the issue, contributes to a learning process, and effectively contributes to ownership of technology development \cite{andersen_stakeholder_2021, ramirez_strategic_2016}. Working with a diverse group of stakeholders can inform scenario-planning in various ways, such as identifying future trends and helping to frame the prioritization in order to ``reduce the number of trends and challenges and identify the most important driving forces'' \cite{andersen_stakeholder_2021}. Participatory foresight approaches are particularly useful in this regard, as they aim to engage a wide range of stakeholders, including lay stakeholders \cite{brey_ethics_2017, nikolova_rise_2014}. Lay stakeholders possess situational knowledge that enriches expert-driven assessments and fills in blind spots in bottom-up impact exploration and mitigation \cite{metcalf_algorithmic_2021, nikolova_rise_2014}. In addition, participants of scenario planning exercises can assist with informing policy making: ``Once alternative futures are mapped out, augmented with narratives, and vetted for quality, stakeholders can be engaged to assess the implications of the scenarios'' \cite{andersen_stakeholder_2021}. Prior empirical studies on the impacts of generative AI have shown that scenario approaches with diverse stakeholders have been fruitful in mapping and evaluating potential governance approaches \cite{kieslich_anticipating_2024, barnett2024simulating, diakopoulos_anticipating_2021, mesmer_auditing_2023}. 

Thematically, we extend our work on the impact of generative AI on the news environment \cite{barnett2024simulating, kieslich_anticipating_2024}. In this field, generative AI has already had a profound impact, with various application potentials ranging from editing and summarizing to the creation of artificial content (for an overview, see \cite{nishal_envisioning_2023, ap_report_genai}). A recent survey of media leaders indicated that 87 percent of respondents think that newsrooms will be somewhat or completely transformed by generative AI \cite{newman_journalism_2025}. While the implementation of generative AI brings many potential benefits, negative impacts also arise such as problems with accuracy (``hallucinations''), the spread of misinformation, and the loss of autonomy \cite{cools_uses_2024, chu2024misinformation}.

This study extends previous approaches by using scenarios as a basis for identifying stakeholder-action pairs for negative impacts of generative AI in the news environment, i.e., enumerating possible strategies for mitigating the negative impacts outlined in the scenarios and assigning responsibility for them to specific actors that have the potential to mitigate the impact. This approach allows us to gauge lay stakeholders' perceptions and use their expertise to uncover not only what actions they envision, but also who they want to be accountable. This enriches a mostly expert-driven definition of mitigation strategies and enables a more democratic approach to policy development.

\section{Data and Methodology}

\begin{figure}
    \centering\includegraphics[width=0.75\linewidth]{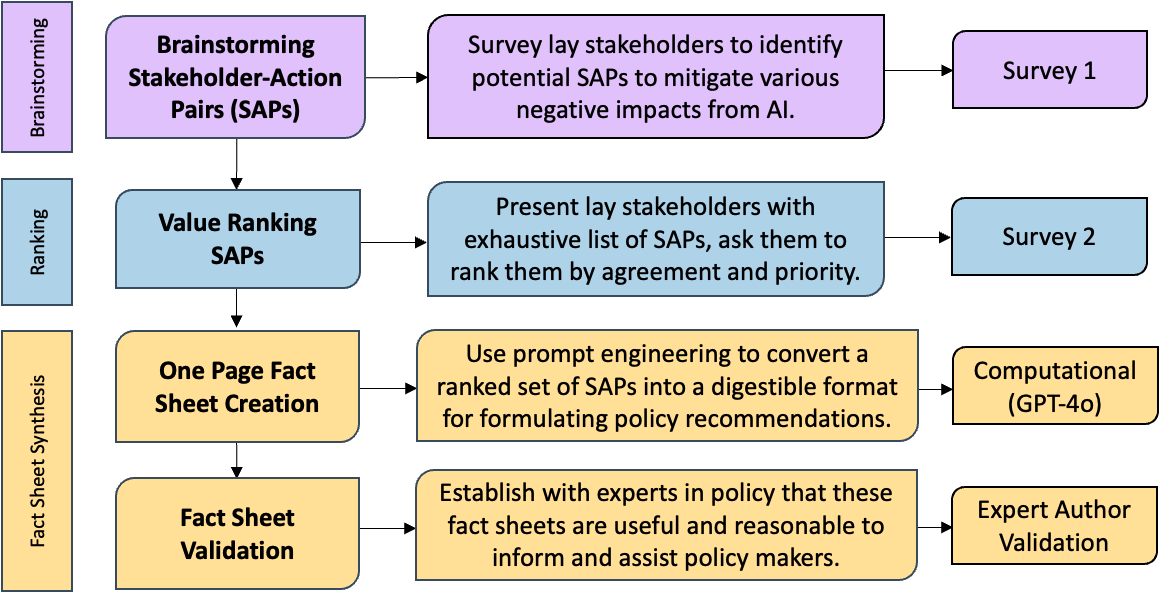}
    \caption{Flow diagram detailing the approach proposed in this work. We first ask lay stakeholders to brainstorm possible stakeholder-action pairs to mitigate negative impacts from AI (Survey 1). We then ask lay stakeholders to rank these in terms of agreement and priority (Survey 2). Finally, we utilize LLMs to synthesize this information into a one-page fact sheet to inform policy makers, and have an expert on our team validate them for usefulness.}
    \label{fig:flow_diagram}
\end{figure}

The goal of this work is to develop a participatory approach to support policy making by using inputs from a broader participative base to uncover the most important needs and prioritizations of lay stakeholders. We did so by first running a survey utilizing written scenarios in order to contextually narrativize the negative impacts for the lay stakeholders to make sure they are comprehensible, then asked them to brainstorm potential responses or interventions to these negative impacts in the form of actions specific actors can take, which we term ``stakeholder-action pairs (SAPs)'' for this work. We then ran a second survey asking lay stakeholders to rank the importance of these SAPs relative to specific impacts in terms of both priority that should be given to this SAP as well as agreement with the proposed SAP. We then used LLMs to convert these results into ``fact-sheets'' synthesizing the most important aspects of these results into one-page guides to inform policy makers. Our full process is illustrated in Figure \ref{fig:flow_diagram}.

\subsection{Selection of Impact Types for Evaluation}

In order to select a set of negative impact types within the broader domain of generative AI in the media, we build on prior work \cite{barnett2024simulating} 
which developed a scenario-based approach to evaluate impacts from a typology 
categorized by impact theme (e.g., media quality) and impact type (e.g., sensationalism). Based on human evaluations of the impact types detailed in the scenarios, that work scored the impact type on a scale from 1 (low) to 5 (high) across four dimensions (severity, plausibility, magnitude, and specificity to vulnerable populations) both pre- and post-introduction of a transparency based policy aimed at mitigating negative impacts from generative AI. We used the pre-policy mitigated rankings of the impacts to identify a smaller set of impacts examined in this work.

More specifically, to identify the set of ten impact types for use in this work, we took a weighted average of severity, plausibility, magnitude, and specificity to vulnerable populations scores with a higher weight (.4) given to severity and an equal weight (.2) given to the other three values. These weights were assigned in order to give greater emphasis to the metric (severity) perceived to have a crucial relevance to risk mitigation prioritization for impact assessment \cite{gellert_risk-based_2020, mesmer_auditing_2023}, while also still taking into account the other dimensions. From this ranked list, one of the authors who is an expert in policy and regulation research evaluated the impact types in regard to the following question: ``Which impact types in this list  are likely receptive to policy intervention?''. This expert evaluation resulted in exchanging three of the top ten impact types from the weighted ranking (\textit{\textbf{Labor}: Changing Job Roles}, \textit{\textbf{Media Quality}: Clickbait}, and \textit{\textbf{Political}: Opinion Monopoly}) with three others from the top 20: \textit{\textbf{Media Quality}}: \textit{Lack of Diversity/Bias}, \textit{Lack of Fact Checking} and \textbf{\textit{Trustworthiness}}: \textit{Overreliance on AI}. Categorized according to the original taxonomy \cite{kieslich_anticipating_2024}.
Our final list of important impact types likely to be receptive to policy intervention is alphabetically as follows: \textit{\textbf{Autonomy}: (1) Loss of Control}, \textit{\textbf{Labor}: (2) Unemployment}, \textit{\textbf{Media Quality}: (3) Credibility/Authenticity, (4) Lack of Diversity/Bias, (5) Lack of Fact Checking, (6) Sensationalism}, \textit{\textbf{Political}: (7) Fake News/Misinformation, (8) Manipulation}, \textit{\textbf{Trustworthiness}: (9) Overreliance on AI}, and (10) \textit{\textbf{Well-Being}: (10) Addiction}. A short description for each impact type can be found in Appendix \ref{sec:appendixA3}.

\subsection{Lay Stakeholder Brainstorming of Approaches to Mitigate Negative Impacts of AI}

In order to be certain lay stakeholders contextually understand the impact types we wish to evaluate in this study, we first provided them with a scenario narrativizing the negative impact in the context of generative AI in the media environment. These are the same scenarios used by \cite{barnett2024simulating}, 
which were written using GPT-4 and created for evaluation of the severity, plausibility, magnitude, and specificity to vulnerable populations that informed the selection of impact types used in this paper. These scenarios are set in the United States five years in the future\footnote{Full scenarios available at: \url{https://tinyurl.com/raw-scenarios-10}}. We decided to use these generated scenarios as opposed to human-written scenarios because LLMs help us create salient narratives to elicit targeted responses whereas human written scenarios tend to be more complex, often intertwining several impact types rather than isolating a specific impact of interest. As such, the LLM-written scenarios tend to be more easily understood and therefore more useful as a tool to get participatory feedback. 

We presented the scenarios to the evaluators, directing them only to pay attention to the specific impact type for which we are eliciting brainstormed SAPs. After they read the scenario narrativizing the impact type, we then asked them to brainstorm four SAPs in the form of stakeholder-action pairs (SAPs). We ask them: ``Specific to harms from generative AI concerning \textit{<\textbf{impact type}>} in regards to <\textit{\textbf{impact theme}}> (e.g., \textit{\textbf{credibility/authenticity}} in regards to \textit{\textbf{media quality}}): What people, organizations, or entities (in real life, not characters in this scenario) do you think could take action in order to prevent the harms illustrated in this scenario from occurring?''. We then provided them four side-by-side free text response boxes where they could fill in the ``People/Organizations/Entities'' and then the corresponding ``Actions they should take.'' They each completed this process three times, seeing one of three scenarios for three randomly assigned impact types, resulting in 12 SAPs brainstormed per participant.

After aggregating scenarios based on impact type, each impact type on average received 48 brainstormed SAPs. We then qualitatively evaluated the proposed stakeholder actions, setting more standardized values for the stakeholders (e.g., ``Congress'' and ``Federal Government'' both became ``Government'') and combining duplicates or similar SAPs for analysis in the second study. 
Evaluators were often more granular in their stipulation of stakeholders---they often described specific branches of government (e.g., legislators or judicial), named specific roles in news publishers (e.g., journalists or editors), and sometimes even mentioned technology companies by name (e.g., Meta or Twitter) or called them ``AI companies,'' but for the purposes of this evaluation the authors qualitatively rolled those into aggregate categories. We also removed SAPs that would not take the form of a policy intervention, such as ``Families and friends need to hold interventions for those addicted to consuming content generated by AI,'' though we do examine these further in our analysis below. We aggregated the proposed SAPs by impact type; after cleaning the SAPs, each impact type had on average 23 SAPs (Median = 22), with a minimum of 16 and maximum of 31, and an overall total of 228. 

\subsection{Establishing Value-Ranking: Agreement, Prioritization, and Consensus}\label{sec:method-value-rank}

With a long unordered list of lay stakeholder generated potential SAPs for these complex impacts, we needed to develop a way to evaluate their relative importance in terms of both agreement the approach should be pursued and the priority with which to do so. For each of the ten impact types, we converted all of the SAPs to ``should'' statements, e.g., ``Schools should promote healthy behavior of generative AI consumption.'' 
We asked them to evaluate a set of SAP ``should'' statements based on \textbf{agreement} using a Likert scale from  1 (Strongly Disagree) to 7 (Strongly Agree) and \textbf{priority} using a ternary scale of Low, Medium, and High. 

We utilized the same set of LLM-generated scenarios as detailed in Survey 1 to be consistent and confident the proposed SAPs will have relevance to the scenarios narrativizing the negative impacts presented to participants. We first surfaced the scenarios to evaluators, again directing them to only pay attention to the specific impact for which we ask them to evaluate. Then we asked them specific to that impact to ``rank how much you agree with the following statements as well as what level of priority you think they should have, in terms of preventing or mitigating the harms detailed in the above scenario.'' We ensured that each SAP received at least 5 evaluations from different participants for each scenario. In total, 86 evaluators took part in the study. Each participant evaluated SAPs for three impact types, and due to the average impact type having 23 finalized SAPs and our necessity of obtaining high quality responses, we asked each participant to evaluate a random subset of 14 SAPs in order to mitigate the potential for respondent fatigue. 

To aggregate and analyze these evaluations, we sorted the evaluations by mean priority value to understand the importance that evaluators attribute to various SAPs. We then note the mean agreement as well as the standard deviation of the agreement to understand if an SAP is generally supported or if it is controversial. The average standard deviation of the agreement scores was 1.5, so we designate any score with a standard deviation less than or equal to 1.5 as higher consensus and those above 1.5 as lower consensus. We analyzed the ranked SAPs within each impact type and compared significant trends across various impact types to understand both what people want to see done and to whom they allocate that responsibility (i.e., which stakeholder they name to take a specific action).


\subsection{Study Recruitment and Participants}

For both of these studies we utilized the research platform Prolific. As the scenarios were located within a US-context, we recruited participants based in the US, fluent in English, with an approval rating of at least 95\%, and at least 100 previous submissions on Prolific. For both studies, each evaluator read three scenarios and answered a short set of questions following each. We ensured the participants received a fair wage for their participation, paying \$4.50 for completion of the first survey (median completion time of 17 minutes) and \$3.25 for completion of the second study (median completion time of 12 minutes), which resulted in an estimated payment of \$15.88/hour and \$16.25/hour respectively. 
We excluded 1 participant in each study for not passing attention checks phrased as nonsensical questions.

For the first study we had a final pool of 40 participants, with self reported demographics of 50\% Male, 40\% Female, and 10\% not consenting to report; 67\% White, 12\% Mixed, 7\% Black, 7\% Asian, and 7\% not consenting to report; median age of 36, with the youngest being 20 and the oldest 61. For the second study, we had a final pool of 86 participants, with self reported demographics of 62\% Female, 36\% Male, 2\% not consenting to report; 59\% White, 12\% Mixed, 9\% Black, 6\% Asian, 10\% Other, and 3\% not consenting to report; median age of 42, with the youngest being 18 and the oldest being 73. We emphasize that this was not a representative sample, and we did not take any steps to ensure equal stratification across any demographic lines.

\subsection{Converting to ``Policy Fact Sheets''}\label{sec:method_policy_fact_sheet}

In order to allow this method to be useful to inform policy-making, it would be helpful to deliver this content in a digestible way to policy-makers. We do this translation step to a concise summary of the survey findings in creating short ``fact sheets'' for each impact type that highlight the high-level trends and aspects lay stakeholders find most useful, while also highlighting aspects that have a lower consensus and might therefore be more contentious. These fact sheets are a proof of concept to illustrate how we could use these findings to craft input for policy makers, however actual documents to inform their decision making would need to incorporate results from a much larger and representative sample of their own communities and jurisdictions. 
To create a streamlined process to create these, we use an LLM (specifically GPT-4o which pointed to ``gpt-4o-2024-08-06'') and prompt engineering to convert this data to a synthesized document of high level trends. We elucidate findings from designing this process with guidance from a member of our team who is an expert in policy and regulation research. Full prompting details can be found in the Appendix \ref{sec:appendix_prompt_engineering}.

\section{Results}

\subsection{Brainstormed Stakeholder-Action Pairs}

We first examine the various SAPs proposed by lay stakeholders from our first survey. After consolidating these as discussed in Section \ref{sec:method-value-rank}, we are left with 228 policy-actionable SAPs across the 10 impact types. After cleaning and consolidating these 228 SAPs, we identified 12 different stakeholders to which responsibility was allocated (see Appendix \ref{sec:stakeholder_definitions} for more detailed definitions of these stakeholders) and 41 different types of actionable approaches encapsulating the various actions specified in the survey. We now examine these by responsibility allocation, actions, the various interactions in which lay stakeholders assigned responsibility to actions, and finally non-policy actionable SAPs.

\begin{figure}
    \centering
\includegraphics[width=0.80\linewidth]{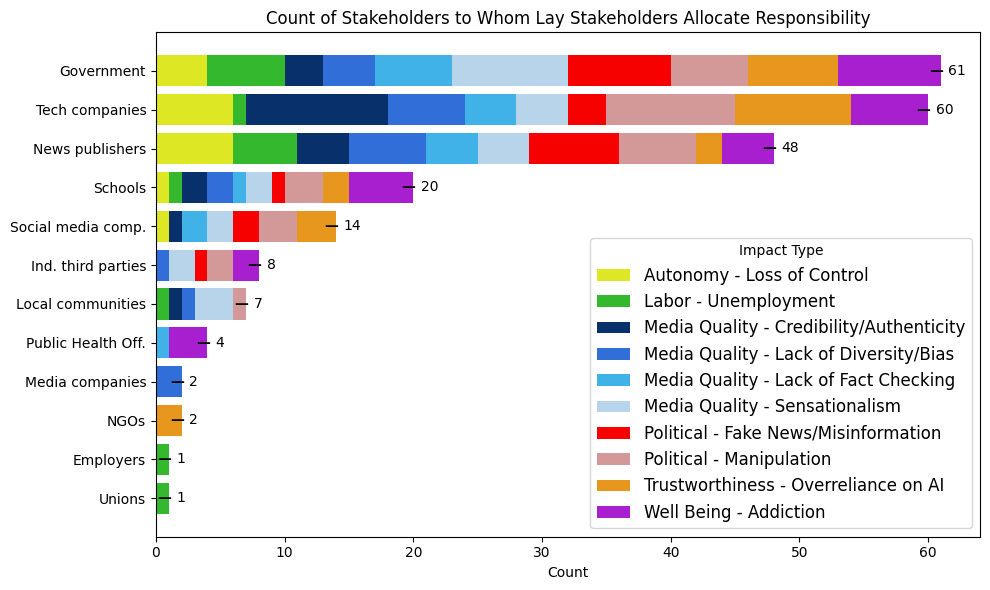}
    \caption{Stacked bar chart displaying how often lay stakeholders allocate responsibility to various stakeholders to take an action, sorted by frequency, split by impact types.}
    \vspace{-3mm}
\label{fig:stacked_bar_stakeholders}
\end{figure}

\textit{\textbf{Responsibility Allocation.}} Across the 10 impact types, we identified 12 different types of stakeholders to whom evaluators assigned responsibility for impact mitigation or prevention, displayed in Figure \ref{fig:stacked_bar_stakeholders}. Of these, government was the most common actor (27\%), followed closely by technology companies (26\%) and news publishers (21\%). 
These three categories appeared in every single impact type, though notably technology companies were allocated responsibility in 50\% of the potential SAPs for credibility/authenticity in regards to media quality, and over a third of the SAPs in both overreliance on AI and loss of control. Government was allocated responsibility in over a third of the SAPs for negative impacts related to unemployment, lack of fact checking, sensationalism, and fake news/misinformation.

For some impact types, respondents allocated responsibility to these three actors the vast majority of the time, though some were much more dispersed. Schools (9\%) were allocated responsibility at least once in each impact type, and social media companies (6\%) were mentioned more often in negative impacts relating to overreliance on AI, fake news/misinformation and manipulation. Proposed responses to negative impacts related to addiction tended to allocate responsibility to a more diverse set of stakeholders than the big three actors (i.e., not just government, tech companies and news publishers); for this impact type responsibility was frequently allocated to schools (18\%), public health officials (11\%), and independent third parties/researchers (7\%) in addition to the big three. Approaches proposed for negative impacts relating to unemployment were allocated to stakeholders such as employers and unions, actors that did not come up in any other category. Local communities (3\%) came up in half of the impact types, but was most commonly discussed in responses to negative impacts relating to sensationalism.

\textit{\textbf{Actionable Approaches.}} After consolidating the SAPs, the various actions proposed in the 228 responses were distilled into 41 distinct actions, which can be found displayed in Figure \ref{fig:stacked_bar_actions} in Appendix \ref{sec:appendix_full__actions}. The most commonly proposed approach was to fact check content or ensure the accuracy of content generated by AI (12\%). The second most popular was educating people by strengthening their digital literacy and critical engagement (9\%), followed by transparency requirements (7\%). Both banning (7\%) and limiting (6\%) the use of AI in writing news articles came up frequently, as well as emphasizing humans over AI or not replacing humans with AI (6\%) and educating people about the general harms resulting from AI (5\%). There were no other actions that came up in more than 5\% of responses.

With the exception of negative impacts related to unemployment, which had a third dedicated to not replacing humans with AI in various ways, every impact type had at least 12 distinct action responses with the average having 16 different action responses. Addiction impacts had the most diverse set of actions (as well as responsibility allocation), with 20 different actions spanning various actors, the most common of which was limiting who has access to AI models (11\%). Both fact checking and educating people about digital literacy came up in every single impact type, though the latter was most common in impacts relating to sensationalism. 11 of the 41 specific actions only came up once each; four of these occurred in impacts related to overreliance on AI with approaches such as ``ban AI companies from using public figures' likeness in training data'' and ``ensure that all newsrooms who use AI have a license to do so.''

\textit{\textbf{Stakeholder-Action Pairs.}}  As a next step, we looked at the interaction of the two aforementioned components: what people want done and by whom. Some actions were only assigned responsibility to certain actors, such as any bans of AI in specific industries or certain penalties like punishing the creation of potentially harmful AI only being allocated to the government. The responsibility of others however, such as educating the public about the harms of AI or strengthening digital literacy and critical engagement, were assigned to multiple actors: government, independent third parties, local communities, news publishers, NGOs, public health officials, schools, social media companies, and technology companies. Some actions were highly concentrated in specific actors but had some others crop up, such as news publishers predominantly receiving responsibility allocation for both fact checking/ensuring accuracy and limiting the use of AI in writing news articles; however, the government, technology companies, independent researchers, and social media companies all were proposed to undertake those actions as well.

\begin{figure}
    \centering
    \includegraphics[width=0.93\linewidth]{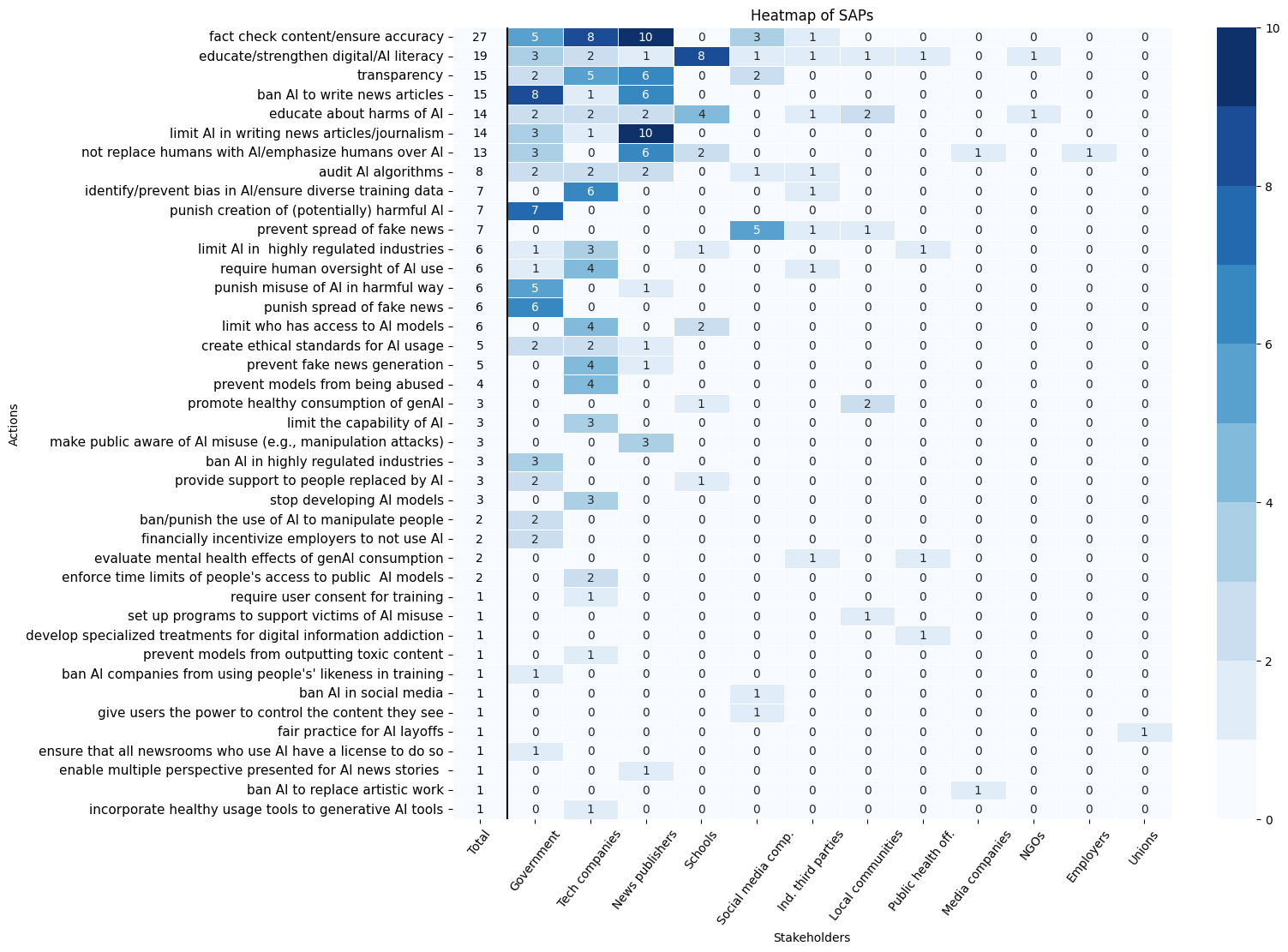}
    \caption{Heatmap displaying the frequency of stakeholder-action pairs across the various impact types brainstormed by lay stakeholders, sorted by vertically by action frequency and horizontally by stakeholder frequency.}
    \label{fig:heat_map}
\end{figure}

Some actions saw an interesting story emerge. For fake news, for example, people proposed multiple actions for both prevention and mitigation across specified actors. Technology companies and news publishers were seen as responsible for preventing the generation of fake news, but social media companies were predominantly seen as the responsible actor in charge of preventing the spread of fake news, and the government was seen as the primary actor in charge of punishing those individuals and entities who (a) create fake news, (b) create technology capable of generating fake news, and (c) spread fake news.

Frequency of suggested SAPs is not necessarily the best metric for evaluating the perceived importance of various approaches---sometimes outliers can be more important than the commonly thought ones. This is why it is essential to not solely evaluate the most frequently brainstormed SAPs, but also to evaluate how lay stakeholders value them. In the next section (Section \ref{sec:results-ranking}), we analyze the SAPs ranked by a set of lay stakeholders to understand which approaches and interventions people actually want to see manifested. 

\textbf{\textit{Non-Policy Implementable Stakeholder-Action Pairs.}} A small subset of SAPs proposed were not implementable by policy, but that does not mean they are not highly relevant to the conversation. Those measures tended to take the form of calling upon individuals' close familial and friendship networks to look after each other. When sensible, we converted some of these types of responses to local communities so that the policy intervention could take the form of allocating financial subsidies supporting these events. However, others were simply omitted from this set such as ``Families encouraging off time and talk[ing] about the major news together.'' 
Future work could examine non-policy initiatives that cater to these types of social and behavioral proposals.

\subsection{Value Ranking the Proposed SAPs}\label{sec:results-ranking}


We next asked evaluators to rank the proposed SAPs (framed as should statements, e.g., ``News publishers should ban AI to write news stories'' by both agreement on a Likert scale of 1 (Strongly Disagree) to 7 (Strongly Agree) and priority (Low, Medium, High). For each impact type, we aggregated and ranked responses by mean priority and then examined the agreement scores by mean and standard deviation. A full ranked list for every impact type can be found in the Appendix \ref{sec:value-ranking} in Tables \ref{tab:Au_LC_results}-\ref{tab:WB_Ad_results}.

On average, our participants thought that most of the brainstormed SAPs should occur. The average agreement score of all SAPs was 5.65/7 (median: 6), and 66\% of evaluations were a 6 or a 7. There was also a consensus that the majority of these items were high priority; when assigned ordinally (Low = 1, Medium = 2, High = 3), the average priority score was 2.28. Overall, 48\% of all items were ranked high priority, 33\% medium priority, and 19\% low priority, indicating predominant support amongst our participants for prioritizing the stakeholder actions we studied. 

Two of the three major stakeholders from the previous section---news publishers and technology companies---still dominate the value ranking. News publishers were assigned responsibility in over a third (36\%) of all SAPs ranked in the top five of the 10 impact types by priority, and technology companies made up one-fifth (22\%). Schools were the next most frequent (18\%), followed by government (10\%). Despite getting allocated responsibility the most out of all SAPs in Survey 1, government proportionally rated unfavorably in terms of the priority ranking falling to fourth place. Almost all of the other stakeholders appeared at least once in the top rankings of SAPs with the exception of NGOs and local communities, both of which consistently rated quite low.

Fact checking persisted as being the highest valued action in terms of priority, with 28\% of the highest rated SAPs having to do with fact checking in some regard, assigned to news publishers (64\%), technology companies (29\%), and social media companies (7\%) in the highest priority rankings. Next came transparency---mostly in terms of labeling something as AI-generated or providing more insight into why models provide given outputs. In these high priority cases this was mostly thought to be the responsibility of news publishers (50\%), tech companies (33\%), and social media companies (17\%). 
Tied for second with transparency in terms of high priority was education. People consistently valued educating both the youth and the greater public about both general harms of AI and how to strengthen digital literacy and critical engagement. Though in the larger list there were many possible stakeholders that people considered allocating responsibility to (schools, government, tech companies, etc.), there was a consensus that lay stakeholders highly valued schools incorporating this into their curriculums rather than other entities taking charge of this.

As we noted earlier, just because recommendations were outliers in terms of how many people came up with the idea in the brainstorming phase did not necessarily make it less important when evaluating all SAPs together. The SAP ``Tech companies should require user consent for training of personalization algorithms/user facing AI,'' only came up once during the brainstorm phase, but it was one of the highest priority actions evaluated within impacts relating to loss of control in autonomy, with a high consensus of people agreeing this should occur ($mean = 6.24; sd = 1.09)$. Similarly, three highly rated unemployment SAPs never came up anywhere else, such as ``Unions should force employers to have fair practice for lay-offs when AI replacements are involved.'' 

Technology companies sometimes received responsibility over consulting another stakeholder, such as ``tech companies should hire independent third parties to monitor and fact check the outputs of their models,'' which indicates that though lay stakeholders believe they should not be the fact checkers themselves, the financial burden of seeing it manifest should still fall on them.

Just as interesting as the most commonly valued SAPs are those that consistently rank among the bottom. The most common two words by far in the bottom ranking of these by both priority and agreement were ``ban'' and ``limit.'' 
Lay stakeholders consistently evaluated outright bans and major limitations on AI or other actors as low priority and low agreement scores. Actions that had responsibility allocated to local communities also tended to rank low; however, this was mostly due to being ranked as low priority rather than lack of agreement. There was not one instance in which local communities taking any action (e.g., ``organize and promote annual get togethers to promote responsible use of AI sources'') ranked outside of the bottom five SAPs.
Some SAPs had high priority but low consensus in terms of agreement. This may indicate some highly contentious SAPs. 
One example of this is within employment: ``News publishers should limit the use of AI in journalism''---this SAP had a mean priority score of 2.62/3, indicating a relatively high priority, but a mean agreement score of 5.54/7 and standard deviation of 2.22 indicating a lower consensus in ratings. Another SAP that had similar scores was ``Employers should not replace human jobs with AI'' (mean priority: 2.25; mean agreement: 5.08; sd: 2.23). People agree these are important discussions to have, but there is not a clear consensus of agreement that those stakeholders should indeed take those actions.


\vspace{-2mm}
\subsection{Policy Fact Sheets}

As discussed in the Methods Section \ref{sec:method_policy_fact_sheet}, as a proof of concept we converted the findings of our surveys into short ``policy fact sheets'' to show how these insights could be delivered to policy makers. A member of our team who is an expert in policy and regulation research helped guide the content we wanted to surface on these documents. We learned through the process that the most useful manner to structure these documents was by presenting policy makers with a whole network of agents that lay stakeholders identify as having a share in ensuring responsible use. As such, we organized the results by the actors (stakeholders) and the actions identified to be taken by them. We also highlight actions by priority and consensus to help elucidate areas in which policy makers may want to tread lightly due to diverging agreement and items that tend to have more widespread support. 

Through presenting the findings in this manner, it became clear that some stakeholder actions were implausible, such as ``use more fact checkers'' or ``automate fact checking,'' either for pragmatic resources reasons or technical limitations.  Some were also politically fraught (and potentially dangerous), such as suggesting governments penalize the production and sharing of false information. This helped guide the structure of these sheets as presentations of what lay stakeholders find useful/not-useful and desirable/undesirable to then be taken into account in a larger discussion of policy development, rather than as direct policy implementations. We also emphasize that when providing this information to policy makers, it is essential to include information about the sample surveyed as well as the use of LLMs in constructing the documents. We present one example fact sheet in the Appendix \ref{sec:appendix_example_fact_sheet} as well as a full repository of them  here: \url{https://tinyurl.com/policy-fact-sheet}.



\vspace{-3mm}
\section{Discussion}

In this work, we developed a participatory approach to support policy making that successfully leverages insights from lay stakeholders---a subsample of the public with no formal or professional link to the technology at hand who nonetheless possess an important kind of situated knowledge. We streamline a process that first (1) narrativizes negative impacts in order for lay stakeholders to brainstorm mitigation and prevention approaches in the form of stakeholder-action pairs, (2) queries their inputs for prioritization and importance, and finally (3) utilizes LLMs to transform this information into a one-page guide to inform policy makers.

\vspace{-3mm}
\subsection{Lay Stakeholders' Value in Mapping Stakeholder-Action Pairs} 

The potential negative impacts of generative AI on the news environment are manifold---so are the ways to mitigate them. Our results show that, also in the eyes of the public, mitigation strategies are neither simple nor tied to one actor. Rather, the respondents of Survey 1 identified a plethora of different SAPs---23 on average per impact type. Overall, we identified 12 different stakeholder groups with 41 different types of actions. This clearly shows that lay people perceptions, i.e., participatory foresight practice, can be fruitful in mapping the field. Further, it supports the usefulness of written scenarios in engaging lay stakeholders in risk mitigation processes \cite{andersen_stakeholder_2021}. 

While there is an emphasis on commonly mentioned actors (governments, news publishers, technology companies), we highlight the interesting finding that some respondents shifted accountability to actors which are mostly neglected in current regulatory debates or impact assessments: schools, for instance, were frequently mentioned as a responsible actor with the primary task to strengthen AI literacy. In the context of unemployment, unions were mentioned as an actor that could actively mitigate negative impacts. Furthermore, local communities and third parties were occasionally mentioned as responsible actors. These responsibility allocations go beyond current regulatory approaches (e.g., the EU AI Act) that normally narrow responsibility down to model providers, system providers, or deployers (e.g., media companies) \cite{carnat_addressing_2024-1} with government bodies as an actor to initiate regulation in the first place. Interestingly, some of the actors mentioned by lay people are in line with the demands of critical stakeholders. For instance, Hartmann and colleagues argue for a strengthening of third parties in auditing \cite{hartmann_addressing_2024}. Also, the frequent mention of educational programs is of interest. While education is identified as one investment area by political decision-makers and in national AI strategies, the kind of literacy that needs to be fostered to counter the harms discussed in this paper are different from the ones that are currently invested in \cite{schiff2022education, foffano2023investing}. As Schiff outlines, citizens need ``Education for AI, \emph{not} AI for Education'' \cite{schiff2022education}, i.e., literacy about contextual and/or ethical aspects of AI and not (technical) training in AI skills. Our findings support this notion as we present contextualized harms in the scenarios. Additionally, many actions proposed by lay stakeholders refer to placing a limitation of some form on AI---be it that AI use should be limited in specific industries (e.g., journalism), for specific tasks (e.g., writing news), limited access for students and minors, or simply not replacing humans with AI. This is an interesting finding insofar as these actions are not typically discussed in regulation. However, some of these actions are already discussed by journalistic actors, who approach governance in terms of the need to develop responsible use guidelines which, for instance, protect the human touch of journalism \cite{cools_uses_2024}.

\subsection{The Contested Nature of (Some) Impact Mitigation Actions} 

In a second step, we evaluated the importance of implementing the suggested SAPs. We took this step because solely mapping potential strategies does not mean that proposed methods are agreed upon by a larger share of lay stakeholders. Some SAPs might only be mentioned occasionally in the brainstorming survey, but are highly valued, some might be politically contested, or some SAPs might simply be considered ineffective.

One of the salient findings of our survey was the contested nature of SAPs that involved any form of regulation or limitation. This can be attributed to the contextual and political setting of the study in the United States since the political approach to AI relies on self rather than state regulation. Future research should try to replicate this study in other contextual settings, for instance, Europe, where the AI Act was recently enacted and survey studies show less reluctance of citizens to opt for regulation \cite{kieslich_regulating_2024}. 
Some of the approaches to mitigate and prevent negative impacts highly valued by lay stakeholders directly contradict with some of the actions taken by large global players. For instance, fact checking was the highest valued action across multiple impact types in this study, and responsibility was allocated to news publishers, technology companies, and social media companies. This desire of the public is highly relevant when put in contrast with recent social media companies' slashing of fact-checking requirements \cite{knibbs_meta_2025, algorithmwatch_statement_2025}. Further, we found that community engagement was ranked low in terms of priority, despite finding approval in general. That indicates that our respondents did not show faith in the efficacy of communities to facilitate active change. 

Understanding the perceived importance of impact mitigation strategies can inform policy decision makers to help prioritize actions and manage their resources. Combined with the results of Survey 1, this can help in enriching risk assessment practices with lay stakeholder input and, thus, close a significant blindspot in current assessments \cite{metcalf_algorithmic_2021, moss_assembling_2021, gillespie_generative_2024, griffin_what_2024}. However, we encourage future research to conduct more higher-powered survey studies (ideally based on representative surveys) to support our observations as previous studies have shown the critical role of public opinion in the development and acceptance of regulatory measures \cite{kieslich_role_2024, rahwan_society---loop_2018, wenzelburger2024algorithms, zhang2020us}. 

\vspace{-2mm}
\subsection{Informing Policy Makers}

We have also shown that the data we collect from surveys can be translated into policy fact sheets. This proof of concept is a practical contribution that enables researchers to make survey results tangible in an easy-to-understand way. As such, we provide translational work that aims to better connect research findings with guidance documents that are often needed by policymakers. The policy fact sheets provide an overview to understand (1) the range of actors responsible for mitigating harm, (2) the variety of possible actions these actors should take, and (3) flag controversial and/or more neutral actions. The policy fact sheets should be understood as supportive tools that can provide decision-makers with an overview for brainstorming and reflection. We emphasize that decision-makers will need to further elaborate some of these actions, i.e., where they are too vague, in order to translate them into feasible actions. 
 
\vspace{-2.5mm}
\subsection{Limitations}

First and foremost, our respondents sample is not a sample of any specific population. We utilized a set of evaluators without any stratified sampling process to serve as our lay stakeholders---those who, we assume, have no formal or professional link to the technology yet are still impacted by it. A more representative sample of respondents would additionally allow for the study of potential demographic differences in evaluations of stakeholder responsibilities, which could be important for helping navigate the politics of policy proposals. However, a larger sample may also introduce new analytic demands, such as clustering stakeholders and actions, where computational methods may need to be integrated and validated in the method. Furthermore, we emphasize in this work the need to involve stakeholders who are most vulnerable to these impacts---future work needs to ensure representing the needs of those vulnerable communities that are disparately impacted in addition to a general subsample of lay stakeholders. 

Despite being helpful in mapping a broad range of SAPs, our method does not allow for a more in-depth and nuanced analysis of the policy suggestions. As a result, the recommendations are often still ``raw'', unnuanced, and not tested upon their compatibility with existing laws and fundamental rights. Insofar, the added value of the method is more on the level of brainstorming, rather than automatically generating ready-to-go policy recommendations.

Further, we have not validated the usefulness of these policy fact sheets with policy makers---we rely instead upon the expertise of one of the authors to guide the usefulness of these documents. Given the breadth of information, the validation by only one expert is also a limitation. Further work needs to examine how best to integrate this information into the actual workflow of policy makers. Additionally, we only evaluate this method in the context of AI governance for negative impacts from generative AI in the media environment. We believe this method would be useful for other emerging technologies or contexts of AI use; however, future work would need to validate the efficacy of this approach in different domains.


\vspace{-3mm}
\section{Conclusion}

In this work, we developed an inclusive approach to inform policy making that utilizes inputs from a broader participative base to create policy suggestions that cover the most important needs and prioritization of lay stakeholders---those who are impacted yet possess no formal or professional link to the technology at hand. We applied the method to negative impacts of generative AI in the media environment to examine and analyze the proposed measures from lay stakeholders to understand where and how a broader participative base believes responsibility should lie in order to mitigate these negative impacts. We then showed how to use LLMs to synthesize this information into a digestible format for policy makers which was validated by an expert in policy on our team. This work supports a vital societal need to integrate more diverse voices in policy making to reflect the needs of those impacted.

\begin{acks}
The funding for this research was provided by UL Research Institutes through the Center for Advancing Safety of Machine
Intelligence. We would also like to thank Mackenzie Jorgensen for her helpful feedback.
\end{acks}

\section*{Author Positionality Statement}
All authors declare that they have no conflicts of interest.

We would also like to address the positionality of the authors. This is an interdisciplinary team of researchers. Julia is a Ph.D. candidate at Northwestern University pursuing a joint doctoral degree in computer science and communications. Her research focuses on algorithmic impact and AI ethics. Kimon is a postdoctoral researcher at the Institute for Information Law at the University of Amsterdam and defines himself as a social scientist. Natali is a Professor of Law and Digital Technology with a special focus on AI. She also holds an honorary doctorate in communication science. Nick is a Professor in Communication Studies and Computer Science (by courtesy) at Northwestern University. His work focuses on computational journalism and AI ethics.

In the current work, Natali served as the primary expert to validate the policy fact sheet, bringing in her academic expertise in the field as well as her experience in serving in diverse roles on expert councils/committees (European Commission, European Parliament, Council of Europe, and national regulators and civil society organizations).

Two of the authors identify as female and two as male. In addition, two authors have a European background and two have a US background. All authors identify as white. We therefore acknowledge a bias in the conduct of this research, as our research team does not reflect the experiences of non-white and non-binary people. We also recognize we all come from countries with a focus on democratic governance that emphasize participatory decision-making, and this has certainly shaped the way we conducted this project.

\section*{Ethical Considerations and Adverse Impacts Statement}

Both studies conducted in this work have been approved (and determined to be exempt) under Institutional Review Board at the host university of the first and last author (Northwestern University). There was also considerable thought and effort made to ensure crowdworkers were paid wages to reflect a living wage. Participants at any time had the chance to withdraw from the study without any penalty. However, we acknowledge that ethical crowdsourcing goes beyond fair pay \cite{shmueli2021beyond, schlagwein2019ethical}, and tested both studies thoroughly prior to launch to be certain there would be no burden to crowdworkers beyond potential discomfort with the harms detailed in the scenarios.

We also recognize that some of the policy recommendations made by lay stakeholders could be potentially dangerous if implemented without nuance or thorough alteration by expert policy makers. One example of this is recommending governments to penalize the production and sharing of false information---this could lead to mass censorship and manipulation of information as determined by the party in power. The goal of this work was to present non-expert biased recommendations by lay stakeholders, so we did not alter or soften any of these suggested approaches, even knowing that some such as this could be potentially dangerous. As such, we do not specifically endorse any of the individual recommendations, but rather the approach as a whole to collect this information.

\bibliographystyle{ACM-Reference-Format}
\bibliography{references}
\clearpage

\appendix
\setcounter{secnumdepth}{2}
\section{Appendix}

\subsection{Stakeholder Definitions}\label{sec:stakeholder_definitions}

Below is an alphabetized list of the stakeholders used in this analysis. This resulted from cleaning the data in Survey 1 where respondents specified stakeholders to take actions to prevent or mitigate the harms of generative AI in the media ecosystem. We then used this cleaned description when mapping the SAPs to the Survey 2.

\begin{itemize}
    \item \textbf{Employers} -- places of work; companies that employ people; named when taking actions over employees in general.
    \item \textbf{Government} -- all branches of government; respondents most frequently used the general term ``government''; designated which branch of government with ``Congress,'' ``president,'' and ``Supreme Court,''; made other specifications such as ``governors,'' ``lawmakers,'' ``policymakers,'' ``regulatory bodies,'' ``regulatory agencies,'' and other one-offs such as a ``government delegated task force.'' 
    \item \textbf{Independent Third Parties} -- typically academics or third party researchers; sometimes more specific such as ``independent fact checkers,'' ``scientists,'' or ``academic AI research community.''
    \item \textbf{Local Communities} -- communities that can organize events or information for a local group of people such as a city or a neighborhood.
    \item \textbf{Media Companies} -- companies specifically named to take action in relation to AI and artistic work; also described as entertainment companies. 
    \item \textbf{News Publishers} -- the most diverse category of stakeholders; includes all named entities or employees relating to news organizations; most frequently referred to as ``news publishers'' or ``news companies''; sometimes specified individual news publishers such as  ``CNN,'' ``Fox News,'' or ``The New York Times''; sometimes specified employee positions such as ``editors,'' ``journalists,'' ``news CEOs,'' or ``news owners.''
    \item \textbf{NGOs} -- non-governmental organizations either specified broadly as ``NGOs'' or specifically as ``charities'' or ``human rights groups.''
    \item \textbf{Public Health Officials} -- this group broadly covers those actually in roles of public health (most commonly attributed in this category) as well as other specified actors within the healthcare domain such as ``doctors,'' ``psychologists,'' and the ``mental health community.''
    \item \textbf{Schools} -- rarely specified which level of schooling; often covering the broad spectrum of K-12 and university level education.
    \item \textbf{Social Media Companies} -- consistently described as ``social media,'' ''social media companies,'' or ``social media sites''; these stakeholders were designated actions specifically dealing with social media platforms. For the purpose of our analysis this did not overlap with \textbf{technology companies.}
    \item \textbf{Technology Companies} -- designated to typically refer to ``Big Tech''; occasionally named entities such as ``OpenAI,'' ``Microsoft,'' or ``Google''; sometimes referred to as  ``AI developers'' or ``AI providers''; infrequently referred to specific roles such as ``AI programmers'' or ``Tech CEOs.'' Distinction was made for \textbf{social media companies} when the specified actions were specifically related to social media.
    \item \textbf{Unions} -- labor unions; named to take action on behalf of workers.
\end{itemize}
\clearpage

\subsection{Full List of Actions Brainstormed}\label{sec:appendix_full__actions}

\begin{figure}[h]
    \centering \includegraphics[width=0.99\linewidth]{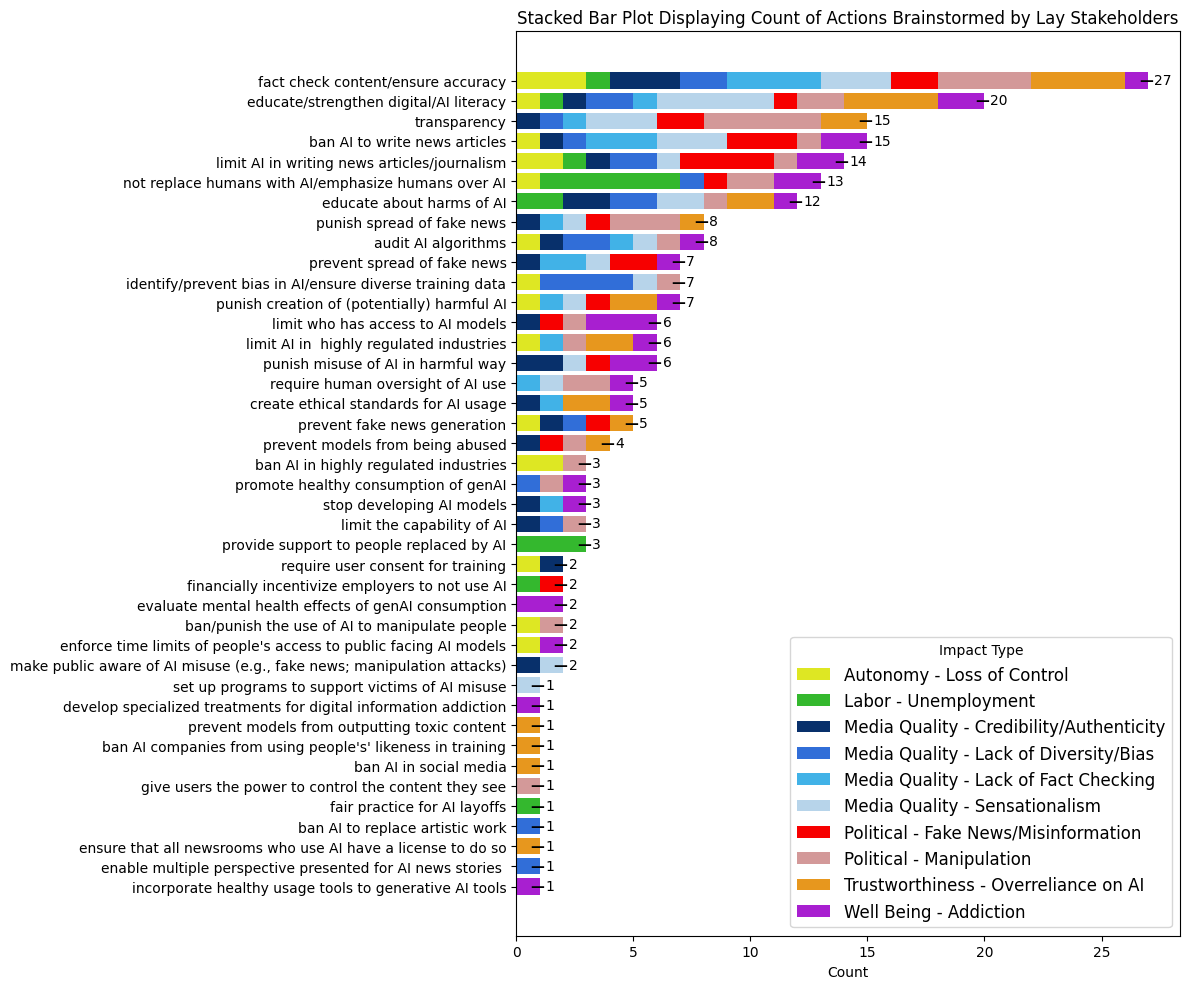}
    \caption{Stacked bar chart displaying how often lay stakeholders brainstormed various actions they wanted stakeholders to take, sorted by frequency, split by impact type.}
    \label{fig:stacked_bar_actions}
\end{figure}
\clearpage

\subsection{Ranked Stakeholder-Action Pair Solutions}
\label{sec:value-ranking}

The following 10 tables display the full ranked results from Survey 2 for each impact type. Results are sorted first by mean priority score (ternary scale of Low=1, Medium=2, High=3), mean agreement score (Likert scale from 1 (Strongly Disagree) to 7 (Strongly Agree), standard deviation of agreement designated to convey consensus. Also included categorical labels for Agreement/Consensus where mean agreement $> 6$ = Higher, $> 3$ = Medium, and $\leq3$ = Lower; Agreement deviation  $\geq1.5$ = Higher, and $<1.5$ = Lower.

\begin{table*}[h]
\begin{small}
\addtolength{\tabcolsep}{-0.08em}
\begin{tabular}{c|c|l|c|c|c|c}\toprule
 \multicolumn{7}{c}{\textbf{Autonomy: Loss of Control} -- Stakeholder-Action Pair Ranking} \\
 \midrule
 \multirow{2}{*}{\textbf{}} &
 \multirow{2}{*}{\textbf{Stakeholder}} &
 \multicolumn{1}{c|}{\multirow{2}{*}{\textbf{Action}}} &
 \textbf{Priority} &
 \textbf{Agree} &
 \textbf{Consensus} &
 \textbf{Agreement/} \\
  &
  &
  &
 (Mean $\uparrow$) &
 (Mean $\uparrow$) &
 (Agree. SD $\downarrow$) &
 \textbf{Consensus} \\
 \cmidrule(lr){1-7}
1 & News publishers & fact check AI generated content. & 2.77 & 6.77 & 0.60 & Higher/Higher \\ \cmidrule(lr){1-7}
\multirow{2}{*}{2} & 
\multirow{2}{*}{Government} & 
punish/fine the companies that create and  & 
\multirow{2}{*}{2.56} & 
\multirow{2}{*}{6.56} & 
\multirow{2}{*}{0.89} & 
\multirow{2}{*}{Higher/Higher} \\ 
& & use AI to engage in harmful acts. & & & & \\
\cmidrule(lr){1-7}
3 & Tech companies & fact check content. & 2.56 & 5.94 & 1.29 & Med/Higher \\ \cmidrule(lr){1-7}
\multirow{2}{*}{4} & 
\multirow{2}{*}{Tech companies} & 
require user consent for training of   & 
\multirow{2}{*}{2.53} & 
\multirow{2}{*}{6.24} & 
\multirow{2}{*}{1.09} & 
\multirow{2}{*}{Higher/Higher} \\ 
& & personalization algorithms/user facing AI. & & & & \\
\cmidrule(lr){1-7}
\multirow{2}{*}{5} & 
\multirow{2}{*}{Schools} & 
strengthen digital literacy and critical  & 
\multirow{2}{*}{2.43} & 
\multirow{2}{*}{6.00} & 
\multirow{2}{*}{1.24} & 
\multirow{2}{*}{Higher/Higher} \\ 
& & engagement with AI.   & & & & \\
\cmidrule(lr){1-7}
6 & Tech companies & incorporate bias detection into AI models. & 2.38 & 6.15 & 1.52 & Higher/Lower \\ \cmidrule(lr){1-7}
7 & News publishers & limit the type of news they use AI for. & 2.38 & 5.75 & 1.29 & Med/Higher \\ \cmidrule(lr){1-7}
8 & Tech companies & prevent AI from generating fake news stories. & 2.36 & 6.14 & 1.10 & Higher/Higher \\ \cmidrule(lr){1-7}
\multirow{2}{*}{9} & 
Social media & 
fact check AI generated content in news  & 
\multirow{2}{*}{2.29} & 
\multirow{2}{*}{5.76} & 
\multirow{2}{*}{1.60} & 
\multirow{2}{*}{Med/Lower} \\ 
& companies & feeds. & & & & \\
\cmidrule(lr){1-7}
10 & News publishers & limit the usage of AI in writing news articles. & 2.29 & 5.71 & 1.44 & Med/Higher \\ \cmidrule(lr){1-7}
11 & Government & make it illegal to use AI to manipulate people. & 2.28 & 5.72 & 1.60 & Med/Lower \\ \cmidrule(lr){1-7}
12 & News publishers & strengthen the human element of journalism. & 2.21 & 5.71 & 1.20 & Med/Higher \\ \cmidrule(lr){1-7}
\multirow{2}{*}{13} & 
\multirow{2}{*}{Tech companies} & 
monitor and restrict what their products  & 
\multirow{2}{*}{2.17} & 
\multirow{2}{*}{5.56} & 
\multirow{2}{*}{1.46} & 
\multirow{2}{*}{Med/Higher} \\ 
& & are being used for.  & & & & \\
\cmidrule(lr){1-7}
14 & News publishers & ban usage of AI to generate news articles. & 2.17 & 4.50 & 2.61 & Med/Lower \\ \cmidrule(lr){1-7}
15 & News publishers & audit AI algorithms. & 2.12 & 5.62 & 1.67 & Med/Lower \\ \cmidrule(lr){1-7}
\multirow{2}{*}{16} & 
\multirow{2}{*}{Government} & 
ban use of AI in heavily regulated industries  & 
\multirow{2}{*}{1.88} & 
\multirow{2}{*}{4.88} & 
\multirow{2}{*}{2.23} & 
\multirow{2}{*}{Med/Lower} \\ 
& & (e.g., medicine, legal, etc.).  & & & & \\
\cmidrule(lr){1-7}
17 & Government & ban the use of AI advertising. & 1.83 & 3.94 & 2.48 & Lower/Lower \\ \cmidrule(lr){1-7}
\multirow{2}{*}{18} & 
\multirow{2}{*}{Tech companies} & 
enforce time limits of people's access to public  & 
\multirow{2}{*}{1.59} & 
\multirow{2}{*}{3.47} & 
\multirow{2}{*}{2.27} & 
\multirow{2}{*}{Lower/Lower} \\ 
& & facing AI models (e.g., ChatGPT). & & & & \\
\cmidrule(lr){1-7}
\bottomrule
\end{tabular}
\end{small}
\caption{Results from Survey 2 ranking stakeholder action pairs specific to harms from generative AI concerning loss of control in regards to autonomy.}
\label{tab:Au_LC_results}
\end{table*}
\clearpage

\begin{table*}[t]
\begin{small}
\addtolength{\tabcolsep}{-0.08em}
\begin{tabular}{c|c|l|c|c|c|c}\toprule
 \multicolumn{7}{c}{\textbf{Labor: Unemployment} -- Stakeholder-Action Pair Ranking} \\
 \midrule
 \multirow{2}{*}{\textbf{}} &
 \multirow{2}{*}{\textbf{Stakeholder}} &
 \multicolumn{1}{c|}{\multirow{2}{*}{\textbf{Action}}} &
 \textbf{Priority} &
 \textbf{Agree} &
 \textbf{Consensus} &
 \textbf{Agreement/} \\
  &
  &
  &
 (Mean $\uparrow$) &
 (Mean $\uparrow$) &
 (Agree. SD $\downarrow$) &
 \textbf{Consensus} \\
 \cmidrule(lr){1-7}
1 & News publishers & employ people to fact check AI. & 2.33 & 6.11 & 1.08 & Higher/Higher \\ \cmidrule(lr){1-7}
\multirow{3}{*}{2} & 
\multirow{3}{*}{Unions} & 
force employers to have fair    & 
\multirow{3}{*}{2.29} & 
\multirow{3}{*}{5.59} & 
\multirow{3}{*}{1.77} & 
\multirow{3}{*}{Med/Lower} \\ 
& & practice for lay-offs when AI   & & & & \\
& & replacements are involved.  & & & & \\
\cmidrule(lr){1-7}
\multirow{3}{*}{3} & 
\multirow{3}{*}{Government} & 
provide financial support and    & 
\multirow{3}{*}{2.29} & 
\multirow{3}{*}{5.29} & 
\multirow{3}{*}{2.14} & 
\multirow{3}{*}{Med/Lower} \\ 
& & benefits to individuals who lost   & & & & \\
& & jobs to AI replacements.  & & & & \\
\cmidrule(lr){1-7}
4 & News publishers & not replace journalists with AI. & 2.26 & 5.42 & 1.89 & Med/Lower \\ \cmidrule(lr){1-7}
5 & Employers & not replace human jobs with AI. & 2.25 & 5.08 & 2.23 & Med/Lower \\ \cmidrule(lr){1-7}
\multirow{2}{*}{6} & 
\multirow{2}{*}{News publishers} & 
strengthen the human element in & 
\multirow{2}{*}{2.22} & 
\multirow{2}{*}{5.56} & 
\multirow{2}{*}{1.46} & 
\multirow{2}{*}{Med/Lower} \\ 
& & journalism.  & & & & \\
\cmidrule(lr){1-7}
\multirow{2}{*}{7} & 
\multirow{2}{*}{Government} & 
require companies to keep a higher & 
\multirow{2}{*}{2.13} & 
\multirow{2}{*}{5.20} & 
\multirow{2}{*}{2.18} & 
\multirow{2}{*}{Med/Lower} \\ 
& & percentage of human employees.  & & & & \\
\cmidrule(lr){1-7}
\multirow{2}{*}{8} & 
\multirow{2}{*}{News publishers} & 
raise awareness about the potential  & 
\multirow{2}{*}{2.06} & 
\multirow{2}{*}{5.59} & 
\multirow{2}{*}{1.33} & 
\multirow{2}{*}{Med/Higher} \\ 
& & harms of AI.  & & & & \\
\cmidrule(lr){1-7}
\multirow{2}{*}{9} & 
\multirow{2}{*}{Schools} & 
provide re-training and re-skilling  & 
\multirow{2}{*}{2.06} & 
\multirow{2}{*}{5.59} & 
\multirow{2}{*}{1.66} & 
\multirow{2}{*}{Med/Lower} \\ 
& & programs for displaced workers.  & & & & \\
\cmidrule(lr){1-7}
10 & News publishers & limit the use of AI in news stories. & 2.06 & 5.53 & 1.59 & Med/Lower \\ \cmidrule(lr){1-7}
\multirow{2}{*}{11} & 
\multirow{2}{*}{Government} & 
make it illegal to replace human jobs & 
\multirow{2}{*}{2.00} & 
\multirow{2}{*}{4.18} & 
\multirow{2}{*}{2.46} & 
\multirow{2}{*}{Med/Lower} \\ 
& & with AI.  & & & & \\
\cmidrule(lr){1-7}
\multirow{2}{*}{12} & 
\multirow{2}{*}{Government} & 
provide universal basic income for & 
\multirow{2}{*}{1.94} & 
\multirow{2}{*}{4.33} & 
\multirow{2}{*}{2.25} & 
\multirow{2}{*}{Med/Lower} \\ 
& & those replaced by AI.  & & & & \\
\cmidrule(lr){1-7}
\multirow{3}{*}{13} & 
\multirow{3}{*}{Government} & 
impose an AI tax for the use of AI  & 
\multirow{3}{*}{1.93} & 
\multirow{3}{*}{5.13} & 
\multirow{3}{*}{1.85} & 
\multirow{3}{*}{Med/Lower} \\ 
& & in companies to encourage the    & & & & \\
& & employment of human jobs.  & & & & \\

\cmidrule(lr){1-7}
\multirow{3}{*}{14} & 
\multirow{3}{*}{Government} & 
pass laws requiring a company to  & 
\multirow{3}{*}{1.88} & 
\multirow{3}{*}{4.44} & 
\multirow{3}{*}{2.31} & 
\multirow{3}{*}{Med/Lower} \\ 
& & have a minimum number of people   & & & & \\
& & employed in order to use AI.  & & & & \\
\cmidrule(lr){1-7}
\multirow{2}{*}{15} & 
\multirow{2}{*}{Tech companies} & 
provide classes for people to learn   & 
\multirow{2}{*}{1.71} & 
\multirow{2}{*}{5.06} & 
\multirow{2}{*}{1.85} & 
\multirow{2}{*}{Med/Lower} \\ 
& & about how to use AI effectively.  & & & & \\
\cmidrule(lr){1-7}
\multirow{2}{*}{16} & 
Local & 
hold discussions on effects of AI on & 
\multirow{2}{*}{1.69} & 
\multirow{2}{*}{4.69} & 
\multirow{2}{*}{1.74} & 
\multirow{2}{*}{Med/Lower} \\ 
& communities & unemployment in the community.  & & & & \\
\cmidrule(lr){1-7}
\bottomrule
\end{tabular}
\end{small}
\caption{Results from Survey 2 ranking stakeholder action pairs specific to harms from generative AI concerning unemployment in regards to labor harms.}
\label{tab:La_Un_results}
\end{table*}
\clearpage

\begin{table*}[t]
\begin{small}
\addtolength{\tabcolsep}{-0.08em}
\begin{tabular}{c|c|l|c|c|c|c}\toprule
 \multicolumn{7}{c}{\textbf{Media Quality: Credibility and Authenticity} -- Stakeholder-Action Pair Ranking} \\
 \midrule
 \multirow{2}{*}{\textbf{}} &
 \multirow{2}{*}{\textbf{Stakeholder}} &
 \multicolumn{1}{c|}{\multirow{2}{*}{\textbf{Action}}} &
 \textbf{Priority} &
 \textbf{Agree} &
 \textbf{Consensus} &
 \textbf{Agreement/} \\
  &
  &
  &
 (Mean $\uparrow$) &
 (Mean $\uparrow$) &
 (Agree. SD $\downarrow$) &
 \textbf{Consensus} \\
 \cmidrule(lr){1-7}
 1 & News publishers & fact check all stories that use AI. & 2.89 & 6.68 & 0.58 & Higher/Higher \\ \cmidrule(lr){1-7}
2 & Tech companies & ensure accuracy of outputs in AI. & 2.76 & 6.59 & 0.62 & Higher/Higher \\ \cmidrule(lr){1-7}
3 & News publishers & recant fake news once made aware. & 2.76 & 6.59 & 0.80 & Higher/Higher \\ \cmidrule(lr){1-7}
\multirow{2}{*}{4} & 
\multirow{2}{*}{Tech companies} & 
follow a standard set of ethical principles  & 
\multirow{2}{*}{2.69} & 
\multirow{2}{*}{6.25} & 
\multirow{2}{*}{0.86} & 
\multirow{2}{*}{Higher/Higher} \\ 
& & to prevent harms of AI.  & & & & \\
\cmidrule(lr){1-7}
\multirow{2}{*}{5} & 
\multirow{2}{*}{Government} & 
take legal action against those who  & 
\multirow{2}{*}{2.67} & 
\multirow{2}{*}{6.20} & 
\multirow{2}{*}{1.37} & 
\multirow{2}{*}{Higher/Higher} \\ 
& & misuse AI. & & & & \\
\cmidrule(lr){1-7}
\multirow{2}{*}{6} & 
\multirow{2}{*}{Tech companies} & 
develop systems to prevent the  & 
\multirow{2}{*}{2.60} & 
\multirow{2}{*}{6.40} & 
\multirow{2}{*}{0.91} & 
\multirow{2}{*}{Higher/Higher} \\ 
& & generation of fake news.  & & & & \\
\cmidrule(lr){1-7}
\multirow{2}{*}{7} & 
\multirow{2}{*}{Tech companies} & 
be transparent about when content is  & 
\multirow{2}{*}{2.56} & 
\multirow{2}{*}{6.25} & 
\multirow{2}{*}{ 1.24 } & 
\multirow{2}{*}{Higher/Higher} \\ 
& & generated by AI.  & & & & \\
\cmidrule(lr){1-7}
\multirow{2}{*}{8} & 
\multirow{2}{*}{Schools} & 
educate people on digital and   & 
\multirow{2}{*}{2.55} & 
\multirow{2}{*}{6.45} & 
\multirow{2}{*}{0.69} & 
\multirow{2}{*}{Higher/Higher} \\ 
& & media literacy. & & & & \\\cmidrule(lr){1-7}
\multirow{2}{*}{9} & 
\multirow{2}{*}{Tech companies} & 
impose safeguards in AI models to  & 
\multirow{2}{*}{2.53} & 
\multirow{2}{*}{6.53} & 
\multirow{2}{*}{0.74} & 
\multirow{2}{*}{Higher/Higher} \\ 
& & prevent misuse. & & & & \\\cmidrule(lr){1-7}
\multirow{2}{*}{10} & 
\multirow{2}{*}{Tech companies} & 
incorporate fact-checking measures into  & 
\multirow{2}{*}{2.53} & 
\multirow{2}{*}{6.41} & 
\multirow{2}{*}{0.94} & 
\multirow{2}{*}{Higher/Higher} \\ 
& & their algorithms. & & & & \\
\cmidrule(lr){1-7}
11 & Schools & teach kids about potential harms of AI. & 2.50 & 6.12 & 0.89 & Higher/Higher \\ \cmidrule(lr){1-7}
\multirow{2}{*}{12} & 
Social media & 
stop promoting sensationalized content  & 
\multirow{2}{*}{2.47} & 
\multirow{2}{*}{5.87} & 
\multirow{2}{*}{1.73} & 
\multirow{2}{*}{ Med/Lower} \\ 
& companies &  in newsfeed. & & & & \\
\cmidrule(lr){1-7}
\multirow{2}{*}{13} & 
\multirow{2}{*}{Government} & 
fine news organizations that disseminate & 
\multirow{2}{*}{2.40} & 
\multirow{2}{*}{5.73} & 
\multirow{2}{*}{1.67} & 
\multirow{2}{*}{Med/Lower} \\ 
& &  fake news. & & & & \\\cmidrule(lr){1-7}
14 & News publishers & limit the use of AI in journalism. & 2.38 & 5.56 & 1.67 & Med/Lower \\ \cmidrule(lr){1-7}
\multirow{2}{*}{15} & 
\multirow{2}{*}{Government} & 
make it illegal to misuse AI and  & 
\multirow{2}{*}{2.36} & 
\multirow{2}{*}{5.91} & 
\multirow{2}{*}{1.14} & 
\multirow{2}{*}{Med/Higher} \\ 
& &  enforce punishments. & & & & \\
\cmidrule(lr){1-7}
16 & Tech companies & stop selling personal data to train models. & 2.22 & 5.44 & 1.76 & Med/Lower \\ \cmidrule(lr){1-7}
\multirow{2}{*}{17} & 
Local & 
organize and educate the youth about the   & 
\multirow{2}{*}{2.06} & 
\multirow{2}{*}{5.71} & 
\multirow{2}{*}{1.49} & 
\multirow{2}{*}{Med/Higher} \\ 
& communities & dangers of misinformation. & & & & \\
\cmidrule(lr){1-7}
18 & Tech companies & halt development of AI systems. & 2.00 & 4.23 & 2.13 & Med/Lower \\ \cmidrule(lr){1-7}
19 & Tech companies & limit who has access to AI models. & 1.93 & 4.87 & 2.07 & Med/Lower \\ \cmidrule(lr){1-7}
20 & News publishers & ban the use of AI to write full stories. & 1.86 & 4.50 & 1.91 & Med/Lower \\ \cmidrule(lr){1-7}
21 & Tech companies & audit competitors' AI algorithms. & 1.79 & 5.07 & 1.90 & Med/Lower \\ \cmidrule(lr){1-7}
22 & Tech companies & limit the capability of AI. & 1.71 & 4.29 & 1.82 & Med/Lower \\ \cmidrule(lr){1-7}
\bottomrule
\end{tabular}
\end{small}
\caption{Results from Survey 2 ranking stakeholder action pairs specific to harms from generative AI concerning credibility and authenticity in regards to media quality.}
\label{tab:MQ_CA_results}
\end{table*}
\clearpage
\begin{table*}[t]
\begin{small}
\addtolength{\tabcolsep}{-0.08em}
\begin{tabular}{c|c|l|c|c|c|c}\toprule
 \multicolumn{7}{c}{\textbf{Media Quality: Lack of Diversity/Bias} -- Stakeholder-Action Pair Ranking} \\
 \midrule
 \multirow{2}{*}{\textbf{}} &
 \multirow{2}{*}{\textbf{Stakeholder}} &
 \multicolumn{1}{c|}{\multirow{2}{*}{\textbf{Action}}} &
 \textbf{Priority} &
 \textbf{Agree} &
 \textbf{Consensus} &
 \textbf{Agreement/} \\
  &
  &
  &
 (Mean $\uparrow$) &
 (Mean $\uparrow$) &
 (Agree. SD $\downarrow$) &
 \textbf{Consensus} \\
 \cmidrule(lr){1-7}
1 & News publishers & fact check all stories that use AI. & 2.94 & 6.94 & 0.24 & Higher/Higher \\ \cmidrule(lr){1-7}
\multirow{2}{*}{2} & 
Independent & 
develop methods to identify bias in & 
\multirow{2}{*}{2.77} & 
\multirow{2}{*}{6.38} & 
\multirow{2}{*}{0.77} & 
\multirow{2}{*}{Higher/Higher} \\ 
&third parties & AI models. & & & & \\
\cmidrule(lr){1-7}
3 & Schools & educate people on the harms of AI. & 2.74 & 6.74 & 0.56 & Higher/Higher \\ \cmidrule(lr){1-7}
\multirow{2}{*}{4 } & 
\multirow{2}{*}{News publishers} & 
implement software to identify & 
\multirow{2}{*}{2.67} & 
\multirow{2}{*}{6.40} & 
\multirow{2}{*}{0.74} & 
\multirow{2}{*}{Higher/Higher} \\ 
& & misinformation. &  & & & \\
\cmidrule(lr){1-7}
5 & News publishers & limit the use of AI in journalism. & 2.62 & 5.54 & 2.22 & Med/Lower \\ \cmidrule(lr){1-7}
\multirow{2}{*}{6} & 
\multirow{2}{*}{Media companies} & 
strengthen the human element in & 
\multirow{2}{*}{2.53} & 
\multirow{2}{*}{6.67} & 
\multirow{2}{*}{0.62} & 
\multirow{2}{*}{Higher/Higher} \\ 
& & content production. &  & & & \\
\cmidrule(lr){1-7}
\multirow{2}{*}{7} & 
\multirow{2}{*}{Tech companies} & 
hire independent third parties to    & 
\multirow{2}{*}{2.53} & 
\multirow{2}{*}{6.40} & 
\multirow{2}{*}{1.30} & 
\multirow{2}{*}{Higher/Higher} \\ 
& & check their models for biases. & & & & \\
\cmidrule(lr){1-7}
8 & News publishers & audit algorithms that use AI. & 2.50 & 5.83 & 1.29 & Med/Higher \\ \cmidrule(lr){1-7}
9 & News publishers & ban the use of AI in news articles. & 2.50 & 5.21 & 2.29 & Med/Lower \\ \cmidrule(lr){1-7}
\multirow{2}{*}{10} & 
\multirow{2}{*}{Tech companies} & 
not add non-factual flair to AI   & 
\multirow{2}{*}{2.44} & 
\multirow{2}{*}{6.11} & 
\multirow{2}{*}{1.37} & 
\multirow{2}{*}{Higher/Higher} \\ 
& &  generated news reporting.  & & & & \\\cmidrule(lr){1-7}
\multirow{2}{*}{11} & 
\multirow{2}{*}{Tech companies} & 
ensure diversity of outputs to  & 
\multirow{2}{*}{2.42} & 
\multirow{2}{*}{6.08} & 
\multirow{2}{*}{1.08} & 
\multirow{2}{*}{Higher/Higher} \\ 
& &  prevent biases.  & & & & \\\cmidrule(lr){1-7}
12 & Media companies & ban use of AI to replace artistic work. & 2.37 & 5.79 & 1.84 & Med/Lower \\ \cmidrule(lr){1-7}
\multirow{3}{*}{13} & 
\multirow{3}{*}{Government} & 
educate the public about digital   & 
\multirow{3}{*}{2.36} & 
\multirow{3}{*}{6.29} & 
\multirow{3}{*}{0.83} & 
\multirow{3}{*}{Higher/Higher} \\ 
& &   literacy and how to identify bias  & & & & \\
& &   in AI generated content. & & & & \\
\cmidrule(lr){1-7}
14 & Schools & educate journalists on the risks of AI. & 2.29 & 6.07 & 1.14 & Higher/Higher \\ \cmidrule(lr){1-7}
15 & Tech companies & create AI fact checking software. & 2.29 & 5.57 & 1.70 & Med/Lower \\ \cmidrule(lr){1-7}
16 & Government & mandate transparency in models. & 2.07 & 5.57 & 1.74 & Med/Lower \\ \cmidrule(lr){1-7}
\multirow{3}{*}{17} & 
\multirow{3}{*}{Tech companies} & 
ensure diversity of training data  & 
\multirow{3}{*}{2.06} & 
\multirow{3}{*}{5.69} & 
\multirow{3}{*}{1.35} & 
\multirow{3}{*}{Med/Higher} \\ 
& &  to represent diversity of people   & & & & \\
& &  who will use AI.  & & & & \\
\cmidrule(lr){1-7}
18 & Tech companies & limit the capability of AI models. & 2.00 & 4.54 & 2.18 & Med/Lower \\ \cmidrule(lr){1-7}
\multirow{2}{*}{19} & 
\multirow{2}{*}{Government} & 
fund community events to educate on  & 
\multirow{2}{*}{1.95} & 
\multirow{2}{*}{5.21} & 
\multirow{2}{*}{1.81} & 
\multirow{2}{*}{Med/Lower} \\ 
& & media literacy. & & & & \\
\cmidrule(lr){1-7}
\multirow{3}{*}{20} & 
\multirow{3}{*}{News publishers} & 
enable news stories generated   & 
\multirow{3}{*}{1.86} & 
\multirow{3}{*}{5.00} & 
\multirow{3}{*}{1.88} & 
\multirow{3}{*}{Med/Lower} \\ 
& &  with AI to be presented from  & & & & \\
& & multiple perspectives.   & & & & \\
\cmidrule(lr){1-7}
\multirow{3}{*}{21} & 
 & 
organize and promote annual   & 
\multirow{3}{*}{1.80} & 
\multirow{3}{*}{5.00} & 
\multirow{3}{*}{1.96} & 
\multirow{3}{*}{Lower/Lower} \\ 
& Local & get togethers to promote responsible  & & & & \\
& communities & use of AI sources. & & & & \\
\cmidrule(lr){1-7}
\multirow{2}{*}{22} & 
\multirow{2}{*}{Government} & 
audit AI models before the public  & 
\multirow{2}{*}{1.80} & 
\multirow{2}{*}{3.67} & 
\multirow{2}{*}{1.88} & 
\multirow{2}{*}{Lower/Lower} \\ 
& & has access to them. & & & & \\
\cmidrule(lr){1-7}
\bottomrule
\end{tabular}
\end{small}
\caption{Results from Survey 2 ranking stakeholder action pairs specific to harms from generative AI concerning lack of diversity/bias in regards to media quality.}
\label{tab:MQ_LDB_results}
\end{table*}
\clearpage

\begin{table*}[t]
\begin{small}
\addtolength{\tabcolsep}{-0.08em}
\begin{tabular}{c|c|l|c|c|c|c}\toprule
 \multicolumn{7}{c}{\textbf{Media Quality: Lack of Fact Checking} -- Stakeholder-Action Pair Ranking} \\
 \midrule
 \multirow{2}{*}{\textbf{}} &
 \multirow{2}{*}{\textbf{Stakeholder}} &
 \multicolumn{1}{c|}{\multirow{2}{*}{\textbf{Action}}} &
 \textbf{Priority} &
 \textbf{Agree} &
 \textbf{Consensus} &
 \textbf{Agreement/} \\
  &
  &
  &
 (Mean $\uparrow$) &
 (Mean $\uparrow$) &
 (Agree. SD $\downarrow$) &
 \textbf{Consensus} \\
 \cmidrule(lr){1-7}
 1 & News publishers & employ human fact-checking teams. & 2.79 & 6.64 & 1.08 & Higher/Higher \\ \cmidrule(lr){1-7}
2 & News publishers & require fact checking for stories. & 2.71 & 6.41 & 1.12 & Higher/Higher \\ \cmidrule(lr){1-7}
\multirow{2}{*}{3} & 
\multirow{2}{*}{Schools} & 
educate people on digital literacy and & 
\multirow{2}{*}{2.69} & 
\multirow{2}{*}{6.77} & 
\multirow{2}{*}{0.60} & 
\multirow{2}{*}{Higher/Higher} \\ 
& & critical engagement. & & & & \\\cmidrule(lr){1-7}
\multirow{2}{*}{4} & 
Social media & 
provide warnings on posts linking to   & 
\multirow{2}{*}{2.47} & 
\multirow{2}{*}{6.27} & 
\multirow{2}{*}{1.16} & 
\multirow{2}{*}{Higher/Higher} \\ 
& companies&  sources known to provide false claims. & & & & \\
\cmidrule(lr){1-7}
5 & Tech companies & build better fact checking into AI models. & 2.40 & 6.27 & 1.39 & Higher/Higher \\ \cmidrule(lr){1-7}
\multirow{2}{*}{6} & 
\multirow{2}{*}{News publishers} & 
add labeling for when AI was used to & 
\multirow{2}{*}{2.31} & 
\multirow{2}{*}{6.50} & 
\multirow{2}{*}{0.89} & 
\multirow{2}{*}{Higher/Higher} \\ 
& &contribute to news articles.  & & & & \\\cmidrule(lr){1-7}
\multirow{2}{*}{7} & 
\multirow{2}{*}{Government} & 
punish individuals or companies that use & 
\multirow{2}{*}{2.29} & 
\multirow{2}{*}{5.47} & 
\multirow{2}{*}{1.94} & 
\multirow{2}{*}{Med/Lower} \\ 
& &AI to spread fake news.  & & & & \\\cmidrule(lr){1-7}
\multirow{2}{*}{8} & 
Public health & 
limit the use of AI in making healthcare & 
\multirow{2}{*}{2.29} & 
\multirow{2}{*}{5.41} & 
\multirow{2}{*}{ 2.27} & 
\multirow{2}{*}{Med/Lower} \\ 
&officials &decisions.  & & & & \\\cmidrule(lr){1-7}
\multirow{2}{*}{9} & 
\multirow{2}{*}{Government} & 
set accuracy standards for use of AI & 
\multirow{2}{*}{2.29} & 
\multirow{2}{*}{5.48} & 
\multirow{2}{*}{ 2.06} & 
\multirow{2}{*}{Med/Lower} \\ 
& &  in journalism. & & & & \\\cmidrule(lr){1-7}
\multirow{2}{*}{10} & 
\multirow{2}{*}{Government} & 
punish technology companies that build AI & 
\multirow{2}{*}{2.25} & 
\multirow{2}{*}{4.94} & 
\multirow{2}{*}{2.57} & 
\multirow{2}{*}{Med/Lower} \\ 
& &  models that can be used to spread fake news. & & & & \\\cmidrule(lr){1-7}
\multirow{2}{*}{11} & 
\multirow{2}{*}{Government} & 
require thorough fact checking if a news & 
\multirow{2}{*}{2.24} & 
\multirow{2}{*}{5.35} & 
\multirow{2}{*}{2.21} & 
\multirow{2}{*}{Med/Lower} \\ 
& & publisher uses AI to generate articles. & & & & \\\cmidrule(lr){1-7}
\multirow{2}{*}{12} & 
\multirow{2}{*}{Tech companies} & 
require human verification before & 
\multirow{2}{*}{2.13} & 
\multirow{2}{*}{5.87} & 
\multirow{2}{*}{1.36} & 
\multirow{2}{*}{Med/Higher} \\ 
& & using AI models. & & & & \\\cmidrule(lr){1-7}
\multirow{2}{*}{13} & 
\multirow{2}{*}{Government} & 
audit tech companies' models and their & 
\multirow{2}{*}{2.00} & 
\multirow{2}{*}{5.47} & 
\multirow{2}{*}{1.68} & 
\multirow{2}{*}{Med/Lower} \\ 
& & fact checking policies. & & & & \\\cmidrule(lr){1-7}
\multirow{2}{*}{14} & 
Social media & 
regulate and remove AI generated news & 
\multirow{2}{*}{2.00} & 
\multirow{2}{*}{5.43} & 
\multirow{2}{*}{1.70} & 
\multirow{2}{*}{Med/Lower} \\ 
& companies& stories on newsfeeds. & & & & \\\cmidrule(lr){1-7}
\multirow{2}{*}{15} & 
\multirow{2}{*}{Tech companies} & 
set guardrails and safeguards to prevent AI & 
\multirow{2}{*}{1.94} & 
\multirow{2}{*}{5.17} & 
\multirow{2}{*}{1.89} & 
\multirow{2}{*}{Med/Lower} \\ 
& & models from being used in journalism.  & & & & \\\cmidrule(lr){1-7}
16 & Government & ban the use of AI in journalism. & 1.83 & 4.11 & 2.40 & Med/Lower \\ \cmidrule(lr){1-7}
17 & News publishers & ban the use of AI in news articles. & 1.68 & 4.47 & 2.22 & Med/Lower \\ \cmidrule(lr){1-7}
18 & Tech companies & stop developing new AI systems. & 1.65 & 3.18 & 2.07 & Lower/Lower \\ \cmidrule(lr){1-7}
\bottomrule
\end{tabular}
\end{small}
\caption{Results from Survey 2 ranking stakeholder action pairs specific to harms from generative AI concerning lack of fact checking in regards to media quality.}
\label{tab:MQ_LFC_results}
\end{table*}
\clearpage
\begin{table*}[t]
\begin{small}
\addtolength{\tabcolsep}{-0.08em}
\begin{tabular}{c|c|l|c|c|c|c}\toprule
 \multicolumn{7}{c}{\textbf{Media Quality: Sensationalism} -- Stakeholder-Action Pair Ranking} \\
 \midrule
 \multirow{2}{*}{\textbf{}} &
 \multirow{2}{*}{\textbf{Stakeholder}} &
 \multicolumn{1}{c|}{\multirow{2}{*}{\textbf{Action}}} &
 \textbf{Priority} &
 \textbf{Agree} &
 \textbf{Consensus} &
 \textbf{Agree./} \\
  &
  &
  &
 (Mean $\uparrow$) &
 (Mean $\uparrow$) &
 (Agree. SD $\downarrow$) &
 \textbf{Consen.} \\
 \cmidrule(lr){1-7}
 1 & News publishers & label content generated with the assistance of AI. & 2.80 & 6.67 & 0.82 & H/H \\ \cmidrule(lr){1-7}
2 & Schools & educate students on how to identify fake news. & 2.76 & 6.59 & 0.87 & H/H \\ \cmidrule(lr){1-7}
3 & Social med comp. & fact check AI generated content. & 2.75 & 6.38 & 0.96 & H/H \\ \cmidrule(lr){1-7}
4 & Tech companies & provide more transparency about how outputs. & 2.67 & 6.44 & 0.73 & H/H \\ \cmidrule(lr){1-7}
\multirow{2}{*}{5} & 
\multirow{2}{*}{News publishers} & 
fact check all content and & 
\multirow{2}{*}{2.67} & 
\multirow{2}{*}{6.07} & 
\multirow{2}{*}{2.12} & 
\multirow{2}{*}{H/L} \\ 
& & reject sensationalized stories. & & & & \\\cmidrule(lr){1-7}
\multirow{2}{*}{6} & 
\multirow{2}{*}{Government} & 
require any use of AI in journalism & 
\multirow{2}{*}{2.62} & 
\multirow{2}{*}{6.12} & 
\multirow{2}{*}{1.71} & 
\multirow{2}{*}{H/L} \\ 
& & to be labeled as AI generated. & & & & \\\cmidrule(lr){1-7}
7 & Government & fact check politicized news stories. & 2.57 & 5.79 & 1.63 & M/L \\ \cmidrule(lr){1-7}
\multirow{2}{*}{8} & 
\multirow{2}{*}{Government} & 
raise awareness nationally of the potential harms & 
\multirow{2}{*}{2.50} & 
\multirow{2}{*}{5.92} & 
\multirow{2}{*}{2.31} & 
\multirow{2}{*}{M/L} \\ 
& & of AI generated news stories. & & & & \\\cmidrule(lr){1-7}
\multirow{2}{*}{9} & 
\multirow{2}{*}{News publishers} & 
alert the public if there have been any  & 
\multirow{2}{*}{2.47} & 
\multirow{2}{*}{6.33} & 
\multirow{2}{*}{0.98} & 
\multirow{2}{*}{H/L} \\ 
& & sensationalized news stories as a result of AI. & & & & \\\cmidrule(lr){1-7}
10 & Tech companies & reduce biases in the training data of AI models. & 2.44 & 5.69 & 1.82 & M/L \\ \cmidrule(lr){1-7}
11 & Government & make it illegal to use of AI to misrepresent  facts. & 2.41 & 5.35 & 2.32 & M/L \\ \cmidrule(lr){1-7}
12 & News publishers & ban the use of AI in generating stories. & 2.40 & 5.60 & 2.06 & M/L \\ \cmidrule(lr){1-7}
13 & Tech companies & require human verification for use of AI models. & 2.38 & 5.92 & 1.75 & M/L \\ \cmidrule(lr){1-7}
14 & Social med comp. & audit AI models. & 2.37 & 5.84 & 1.98 & M/L \\ \cmidrule(lr){1-7}
\multirow{2}{*}{15} & 
Independent & 
develop educational programs for the public & 
\multirow{2}{*}{2.31} & 
\multirow{2}{*}{6.31} & 
\multirow{2}{*}{0.95} & 
\multirow{2}{*}{H/H} \\ 
& third parties  & about how to critically assess AI. & & & & \\\cmidrule(lr){1-7}
\multirow{2}{*}{16} & 
\multirow{2}{*}{Schools} & 
teach digital literacy and critical engagement to  & 
\multirow{2}{*}{2.29} & 
\multirow{2}{*}{6.21} & 
\multirow{2}{*}{0.97} & 
\multirow{2}{*}{H/H} \\ 
& & help students identify AI generated content. & & & & \\\cmidrule(lr){1-7}
\multirow{2}{*}{17} & 
\multirow{2}{*}{Tech companies} & 
help strengthen digital literacy of the public & 
\multirow{2}{*}{2.27} & 
\multirow{2}{*}{5.82} & 
\multirow{2}{*}{1.89} & 
\multirow{2}{*}{M/L} \\ 
& &  to help them understand harms of AI. & & & & \\\cmidrule(lr){1-7}
18 & Government & impose fines on news publishers for fake news. & 2.27 & 5.73 & 1.98 & M/L \\ \cmidrule(lr){1-7}
\multirow{2}{*}{19} & 
Local & 
provide programs to educate communities & 
\multirow{2}{*}{2.19} & 
\multirow{2}{*}{5.50} & 
\multirow{2}{*}{1.67} & 
\multirow{2}{*}{M/H} \\ 
& communities & on digital literacy. & & & & \\\cmidrule(lr){1-7}
\multirow{2}{*}{20} & 
Local & 
monitor media affecting local communities and  & 
\multirow{2}{*}{2.19} & 
\multirow{2}{*}{5.44} & 
\multirow{2}{*}{1.67} & 
\multirow{2}{*}{H/H} \\ 
& communities & provide counter-evidence to false claims. & & & & \\\cmidrule(lr){1-7}
21 & Government & limit use of AI in news media. & 2.18 & 5.35 & 2.00 & M/L \\ \cmidrule(lr){1-7}
22 & Government & ban the use of AI in journalism. & 2.12 & 4.35 & 2.23 & M/L \\ \cmidrule(lr){1-7}
23 & Ind. third parties & raise awareness about the potential harms of AI. & 2.11 & 5.78 & 1.22 &M/H \\ \cmidrule(lr){1-7}
\multirow{2}{*}{24} & 
\multirow{2}{*}{Government} & 
make it illegal to release AI models that can  & 
\multirow{2}{*}{2.10} & 
\multirow{2}{*}{4.95} & 
\multirow{2}{*}{2.04} & 
\multirow{2}{*}{M/L} \\ 
& &  be used to make defamatory content. & & & & \\\cmidrule(lr){1-7}
25 & Government & ban use of AI generated news stories. & 2.00 & 4.22 & 2.59 & M/L \\ \cmidrule(lr){1-7}
\multirow{2}{*}{26} & 
Local & 
 set up programs to support victims of & 
\multirow{2}{*}{1.71} & 
\multirow{2}{*}{4.50} & 
\multirow{2}{*}{2.18} & 
\multirow{2}{*}{M/L} \\ 
& communities & sensationalized news.  & & & & \\\cmidrule(lr){1-7}\bottomrule
\end{tabular}
\end{small}
\caption{Results from Survey 2 ranking stakeholder action pairs specific to harms from generative AI concerning sensationalism in regards to media quality.}
\label{tab:MQ_Se_results}
\end{table*}
\clearpage
\begin{table*}[t]
\begin{small}
\addtolength{\tabcolsep}{-0.08em}
\begin{tabular}{c|c|l|c|c|c|c}\toprule
 \multicolumn{7}{c}{\textbf{Political: Fake News/Misinformation} -- Stakeholder-Action Pair Ranking} \\
 \midrule
 \multirow{2}{*}{\textbf{}} &
 \multirow{2}{*}{\textbf{Stakeholder}} &
 \multicolumn{1}{c|}{\multirow{2}{*}{\textbf{Action}}} &
 \textbf{Priority } &
 \textbf{Agree} &
 \textbf{Consensus} &
 \textbf{Agreement/} \\
  &
  &
  &
 (Mean $\uparrow$)&
 (Mean $\uparrow$) &
 (Agree. SD  $\downarrow$) &
 \textbf{Consensus} \\
 \cmidrule(lr){1-7}
 1 & News publishers & fact check all stories. & 2.91 & 6.91 & 0.30 & Higher/Higher \\ \cmidrule(lr){1-7}
2 & News publishers & moderate the use of AI in their news stories. & 2.90 & 6.90 & 0.31 & Higher/Higher \\ \cmidrule(lr){1-7}
\multirow{2}{*}{3} & 
\multirow{2}{*}{Schools} & 
teach students how to identify fake news  & 
\multirow{2}{*}{2.71} & 
\multirow{2}{*}{6.71} & 
\multirow{2}{*}{0.47} & 
\multirow{2}{*}{Higher/Higher} \\ 
& & and misinformation. & & & & \\\cmidrule(lr){1-7}
\multirow{2}{*}{4} & 
\multirow{2}{*}{Government} & 
take legal action against people  & 
\multirow{2}{*}{2.71} & 
\multirow{2}{*}{5.53} & 
\multirow{2}{*}{2.29} & 
\multirow{2}{*}{Med/Lower} \\ 
& & intentionally spreading fake news. & & & & \\\cmidrule(lr){1-7}
5 & Social med. comp. & prevent spread of fake news on platforms. & 2.57 & 5.36 & 1.95 & Med/Lower \\ \cmidrule(lr){1-7}
6 & News publishers & label all content generated with the use of AI. & 2.56 & 6.61 & 0.78 & Higher/Higher \\ \cmidrule(lr){1-7}
7 & Social med. comp. & create algorithms to identify AI generated posts. & 2.50 & 6.36 & 0.74 & Higher/Higher \\ \cmidrule(lr){1-7}
8 & News publishers & inform users when AI was used to write articles. & 2.46 & 6.69 & 0.85 & Higher/Higher \\ \cmidrule(lr){1-7}
\multirow{2}{*}{9} & 
\multirow{2}{*}{Tech companies}  & 
prevent generation of fake news in &
\multirow{2}{*}{2.43} & 
\multirow{2}{*}{5.57} & 
\multirow{2}{*}{2.17} & 
\multirow{2}{*}{Med/Lower} \\ 
&  & their AI models. & & & & \\\cmidrule(lr){1-7}
\multirow{2}{*}{10} & 
Independent & 
\multirow{2}{*}{fact check all news stories that are published.}  & 
\multirow{2}{*}{2.42} & 
\multirow{2}{*}{6.32} & 
\multirow{2}{*}{1.00} & 
\multirow{2}{*}{Higher/Higher} \\ 
& third parties &  & & & & \\\cmidrule(lr){1-7}
\multirow{2}{*}{11} & 
\multirow{2}{*}{Government} & 
subsidize human journalists so news  & 
\multirow{2}{*}{ 2.42} & 
\multirow{2}{*}{4.42} & 
\multirow{2}{*}{2.47} & 
\multirow{2}{*}{Med/Lower} \\ 
& &  publishers don't use AI. & & & & \\\cmidrule(lr){1-7}
12 & Government & take legal action against those who misuse AI. & 2.36 & 5.86 & 1.79 & Med/Lower \\ \cmidrule(lr){1-7}
13 & News publishers & limit the use of AI. & 2.31 & 5.69 & 1.74 & Med/Lower \\ \cmidrule(lr){1-7}
\multirow{2}{*}{14} & 
\multirow{2}{*}{ Tech companies} & 
provide safeguards in AI models to  & 
\multirow{2}{*}{2.31} & 
\multirow{2}{*}{5.31} & 
\multirow{2}{*}{2.24} & 
\multirow{2}{*}{Med/Lower} \\ 
& & prevent misuse. & & & & \\\cmidrule(lr){1-7}
15 & Government & limit the usage of AI in news. & 2.21 & 4.29 & 2.52 & Med/Lower \\ \cmidrule(lr){1-7}
\multirow{2}{*}{16} & 
\multirow{2}{*}{Government} & 
make it illegal for tech companies to release   & 
\multirow{2}{*}{2.17} & 
\multirow{2}{*}{4.17} & 
\multirow{2}{*}{2.72} & 
\multirow{2}{*}{Med/Lower} \\ 
& & models capable of generating fake news. & & & & \\\cmidrule(lr){1-7}
\multirow{2}{*}{17} & 
\multirow{2}{*}{News publishers} & 
employ people to specifically analyze news  & 
\multirow{2}{*}{2.11} & 
\multirow{2}{*}{5.39} & 
\multirow{2}{*}{1.65} & 
\multirow{2}{*}{Med/Lower} \\ 
& & content for use of AI. & & & & \\\cmidrule(lr){1-7}
18 & Government & ban the use of AI in journalism. & 2.08 & 3.85 & 2.34 & Lower/Lower \\ \cmidrule(lr){1-7}
19 & Government & ban the use of AI to write news stories. & 2.00 & 4.00 & 2.39 & Med/Lower \\ \cmidrule(lr){1-7}
\multirow{2}{*}{20} & 
\multirow{2}{*}{Government} & 
make it illegal to use AI to generate  & 
\multirow{2}{*}{ 2.00} & 
\multirow{2}{*}{3.75} & 
\multirow{2}{*}{2.41} & 
\multirow{2}{*}{Lower/Lower} \\ 
& &  political news stories. & & & & \\\cmidrule(lr){1-7}
21 & News publishers & strengthen the human element in journalism. & 1.94 & 6.00 & 1.50 & Higher/Higher \\ \cmidrule(lr){1-7}
22 & Tech companies & limit access to the use of AI models. & 1.59 & 3.41 & 2.29 & Lower/Lower \\ \cmidrule(lr){1-7}

\bottomrule
\end{tabular}
\end{small}
\caption{Results from Survey 2 ranking stakeholder action pairs specific to harms from generative AI concerning fake news/misinformation in regards to political harms.}
\label{tab:Po_FN_results}
\end{table*}
\clearpage
\begin{table*}[t]
\begin{small}
\addtolength{\tabcolsep}{-0.1em}
\begin{tabular}{c|c|l|c|c|c|c}\toprule
 \multicolumn{7}{c}{\textbf{Political: Manipulation} -- Stakeholder-Action Pair Ranking} \\
 \midrule
 \multirow{2}{*}{\textbf{}} &
 \multirow{2}{*}{\textbf{Stakeholder}} &
 \multicolumn{1}{c|}{\multirow{2}{*}{\textbf{Action}}} &
 \textbf{Priority} &
 \textbf{Agree } &
 \textbf{Consensus } &
 \textbf{Agree./} \\
  &
  &
  &
 (Mean $\uparrow$) &
 (Mean $\uparrow$) &
 (Agree. SD $\downarrow$) &
 \textbf{Consen.} \\
 \cmidrule(lr){1-7}
 1 & Tech companies & be transparent about data collection and usage. & 2.82 & 7.00 & 0.00 & H/H \\ \cmidrule(lr){1-7}
2 & News publishers & be transparent about any use of AI in stories. & 2.80 & 6.75 & 0.55 & H/H \\ \cmidrule(lr){1-7}
3 & Tech companies & build safeguards into models to prevent misuse. & 2.80 & 6.60 & 0.63 & H/H \\ \cmidrule(lr){1-7}
\multirow{2}{*}{4} & 
\multirow{2}{*}{News publishers} & 
provide tools to help identify when  & 
\multirow{2}{*}{2.80} & 
\multirow{2}{*}{6.60} & 
\multirow{2}{*}{0.97} & 
\multirow{2}{*}{H/H} \\ 
& & articles are AI generated. & & & & \\\cmidrule(lr){1-7}
5 & Schools & strengthen digital literacy \& critical engagement. & 2.75 & 6.56 & 0.63 & H/H \\ \cmidrule(lr){1-7}
6 & News publishers & fact check all stories. & 2.69 & 6.92 & 0.28 & H/H \\ \cmidrule(lr){1-7}
7 & Social med. comp. & have labels identifying when a post is AI generated. & 2.69 & 6.75 & 0.45 & H/H \\ \cmidrule(lr){1-7}
8 & Government & mandate tech companies \& news pub. fact check. & 2.67 & 6.33 & 1.78 & H/L \\ \cmidrule(lr){1-7}
9 & Tech companies & include cited sources in outputs of models. & 2.57 & 6.64 & 0.63 & H/H \\ \cmidrule(lr){1-7}
10 & Tech companies & limit usage of their AI models in news \& politics. & 2.55 & 6.09 & 1.30 & H/H \\ \cmidrule(lr){1-7}
11 & Schools & educate students about potential harms of AI. & 2.53 & 6.79 & 0.54 & H/H \\ \cmidrule(lr){1-7}
12 & Government & raise awareness about harms of AI. & 2.53 & 6.42 & 1.02 & H/H \\ \cmidrule(lr){1-7}
13 & News publishers & make the public aware of AI manipulation attacks. & 2.50 & 6.50 & 0.85 & H/H \\ \cmidrule(lr){1-7}
14 & News publishers & strengthen the human element in their journalism. & 2.50 & 6.42 & 0.51 & H/H \\ \cmidrule(lr){1-7}
\multirow{2}{*}{18} & 
\multirow{2}{*}{News publishers} & 
strengthen public literacy by highlighting  & 
\multirow{2}{*}{2.47} & 
\multirow{2}{*}{6.20} & 
\multirow{2}{*}{1.08} & 
\multirow{2}{*}{H/H} \\ 
& & identified fake news stories. & & & & \\\cmidrule(lr){1-7}
16 & Tech companies & require human verification prior to using AI models. & 2.42 & 6.26 & 1.10 & H/H \\ \cmidrule(lr){1-7}
17 & Government & regulate news companies' use of AI. & 2.40 & 6.25 & 1.12 & H/H \\ \cmidrule(lr){1-7}
\multirow{2}{*}{18} & 
\multirow{2}{*}{Government} & 
take legal action against news organizations  & 
\multirow{2}{*}{2.40} & 
\multirow{2}{*}{5.80} & 
\multirow{2}{*}{1.47} & 
\multirow{2}{*}{M/H} \\ 
& & spreading fake news. & & & & \\\cmidrule(lr){1-7}
19 & Schools & strengthen human values and interactions over AI. & 2.38 & 5.54 & 1.76 & M/L \\ \cmidrule(lr){1-7}
20 & Social med. comp. & allow users to have more control over newsfeeds. & 2.36 & 6.21 & 1.19 & H/H \\ \cmidrule(lr){1-7}
21 & Government & make it a crime to manipulate someone using AI. & 2.31 & 6.08 & 1.44 & H/H \\ \cmidrule(lr){1-7}
22 & Tech companies & require fact checking before generating false stories. & 2.29 & 6.29 & 1.20 & H/H \\ \cmidrule(lr){1-7}
23 & Tech companies & limit the capabilities of AI. & 2.27 & 5.64 & 1.96 & M/L \\ \cmidrule(lr){1-7}
24 & Tech companies & remove toxicity from data prior to training models. & 2.26 & 6.21 & 0.98 & H/H \\ \cmidrule(lr){1-7}
25 & Ind. third parties & verify that accounts online are created by humans. & 2.25 & 5.67 & 1.50 & M/H \\ \cmidrule(lr){1-7}
26 & Ind. third parties & audit tech companies' AI models. & 2.24 & 6.18 & 0.81 & H/H \\ \cmidrule(lr){1-7}
27 & Tech companies & ban AI usage by news publishers to generate stories. & 2.24 & 5.29 & 1.72 & M/L \\ \cmidrule(lr){1-7}
28 & Social med. comp. & fact check all posts. & 2.08 & 6.08 & 1.00 & H/H \\ \cmidrule(lr){1-7}
29 & Tech companies & limit access to the use of AI models. & 2.05 & 4.80 & 2.12 & M/L \\ \cmidrule(lr){1-7}
30 & Government & ban the use of AI. & 1.75 & 4.06 & 2.54 & M/L \\ \cmidrule(lr){1-7}
\multirow{2}{*}{31} & 
Local & 
organize and promote annual get togethers to  & 
\multirow{2}{*}{1.56} & 
\multirow{2}{*}{5.06} & 
\multirow{2}{*}{1.61} & 
\multirow{2}{*}{M/L} \\ 
& communities & promote responsible use of AI. & & & & \\\cmidrule(lr){1-7}
\bottomrule
\end{tabular}
\end{small}
\caption{Results from Survey 2 ranking stakeholder action pairs specific to harms from generative AI concerning manipulation in regards to political harms.}
\label{tab:Po_Ma_results}
\end{table*}
\clearpage
\begin{table*}[t]
\begin{small}
\addtolength{\tabcolsep}{-0.08em}
\begin{tabular}{c|c|l|c|c|c|c}\toprule
 \multicolumn{7}{c}{\textbf{Trustworthiness: Overreliance on AI} -- Stakeholder-Action Pair Ranking} \\
 \midrule
 \multirow{2}{*}{\textbf{}} &
 \multirow{2}{*}{\textbf{Stakeholder}} &
 \multicolumn{1}{c|}{\multirow{2}{*}{\textbf{Action}}} &
 \textbf{Priority} &
 \textbf{Agree} &
 \textbf{Consensus} &
 \textbf{Agree./} \\
  &
  &
  &
 (Mean $\uparrow$) &
 (Mean $\uparrow$) &
 (Agree. SD $\downarrow$) &
 \textbf{Consen.} \\
 \cmidrule(lr){1-7}
  1 & News publishers & fact check all stories that used AI. & 2.89 & 6.56 & 0.73 & H/H \\ \cmidrule(lr){1-7}
2 & Tech companies & fact check the outputs of their models. & 2.78 & 6.11 & 1.96 & H/L \\ \cmidrule(lr){1-7}
\multirow{2}{*}{3} & 
\multirow{2}{*}{Tech companies} & 
build safeguards into AI models to & 
\multirow{2}{*}{2.71} & 
\multirow{2}{*}{6.59} & 
\multirow{2}{*}{0.71} & 
\multirow{2}{*}{H/H} \\ 
& & prevent fake news generation. & & & & \\\cmidrule(lr){1-7}
\multirow{2}{*}{4} & 
\multirow{2}{*}{Schools} & 
educate people about digital literacy & 
\multirow{2}{*}{2.62} & 
\multirow{2}{*}{6.31} & 
\multirow{2}{*}{1.18} & 
\multirow{2}{*}{H/H} \\ 
& & and critical engagement. & & & & \\\cmidrule(lr){1-7}
5 & Social med comp. & clearly label AI generated content. & 2.58 & 6.47 & 0.84 & H/H \\ \cmidrule(lr){1-7}
\multirow{2}{*}{6} & 
\multirow{2}{*}{Tech companies} & 
be transparent about when outputs of their & 
\multirow{2}{*}{2.53} & 
\multirow{2}{*}{6.47} & 
\multirow{2}{*}{0.87} & 
\multirow{2}{*}{H/H} \\ 
& & models might be misinformation. & & & & \\\cmidrule(lr){1-7}
\multirow{2}{*}{7} & 
\multirow{2}{*}{Tech companies} & 
prevent the ability to impersonate public & 
\multirow{2}{*}{2.53} & 
\multirow{2}{*}{6.68} & 
\multirow{2}{*}{0.48} & 
\multirow{2}{*}{H/H} \\ 
& & figures (deepfakes) in models. & & & & \\\cmidrule(lr){1-7}
8 & Tech companies & strengthen ethical AI development. & 2.50 & 6.29 & 1.07 & H/H \\ \cmidrule(lr){1-7}
\multirow{2}{*}{9} & 
\multirow{2}{*}{Tech companies} & 
hire independent third parties to monitor \&  & 
\multirow{2}{*}{2.50} & 
\multirow{2}{*}{6.14} & 
\multirow{2}{*}{1.17} & 
\multirow{2}{*}{H/H} \\ 
& & fact check the outputs of their models. & & & & \\\cmidrule(lr){1-7}
10 & Schools & monitor use of AI in assignments. & 2.50 & 6.12 & 0.96 & H/H \\ \cmidrule(lr){1-7}
\multirow{1}{*}{11} & 
\multirow{1}{*}{Tech companies} & 
warn the public about the potential harms of AI. & 
\multirow{1}{*}{2.50} & 
\multirow{1}{*}{5.94} & 
\multirow{1}{*}{1.34} & 
\multirow{1}{*}{M/H} \\\cmidrule(lr){1-7}
\multirow{2}{*}{12} & 
\multirow{2}{*}{Government} & 
hold tech companies accountable for the \&  & 
\multirow{2}{*}{2.47} & 
\multirow{2}{*}{6.18} & 
\multirow{2}{*}{1.38} & 
\multirow{2}{*}{H/H} \\ 
& & false information their models produce. & & & & \\\cmidrule(lr){1-7}
\multirow{2}{*}{13} & 
\multirow{2}{*}{News publishers} & 
create and adhere to a set of standards for  & 
\multirow{2}{*}{2.47} & 
\multirow{2}{*}{6.07} & 
\multirow{2}{*}{1.16} & 
\multirow{2}{*}{H/H} \\ 
&  & all ethical usage of AI. & & & & \\\cmidrule(lr){1-7}
14 & Tech companies & prevent their models from making toxic content. & 2.42 & 5.95 & 1.39 & M/H \\ \cmidrule(lr){1-7}
15 & NGOs & educate the public and strengthen digital literacy. & 2.36 & 6.00 & 1.18 & H/H \\ \cmidrule(lr){1-7}
16 & Government & make it illegal to produce fake news. & 2.25 & 5.50 & 1.55 & M/L \\ \cmidrule(lr){1-7}
\multirow{2}{*}{17} & 
Social media & 
remind users to strengthen critical engagement  & 
\multirow{2}{*}{2.20} & 
\multirow{2}{*}{6.00} & 
\multirow{2}{*}{1.36} & 
\multirow{2}{*}{H/H} \\ 
& companies & and not to trust everything they read. & & & & \\\cmidrule(lr){1-7}
\multirow{2}{*}{19} & 
\multirow{2}{*}{Government} & 
ensure that all newsrooms who use AI have & 
\multirow{2}{*}{2.19 } & 
\multirow{2}{*}{5.75} & 
\multirow{2}{*}{1.18} & 
\multirow{2}{*}{M/H} \\ 
& & a license to do so. & & & & \\\cmidrule(lr){1-7}
\multirow{2}{*}{19} & 
\multirow{2}{*}{Government} & 
shut down AI companies that create models  & 
\multirow{2}{*}{2.14 } & 
\multirow{2}{*}{5.07} & 
\multirow{2}{*}{2.20} & 
\multirow{2}{*}{M/L} \\ 
& & capable of generating deepfakes. & & & & \\\cmidrule(lr){1-7}
\multirow{2}{*}{20} & 
\multirow{2}{*}{Government} & 
require news articles to go through a  & 
\multirow{2}{*}{2.07} & 
\multirow{2}{*}{5.36} & 
\multirow{2}{*}{2.24} & 
\multirow{2}{*}{M/L} \\ 
& & fact checking verification process. & & & & \\\cmidrule(lr){1-7}
21 & Tech companies & moderate in which domains their tech can be used. & 2.00 & 5.53 & 1.22 & M/H \\ \cmidrule(lr){1-7}
22 & NGOs & raise awareness of the potential harms of AI. & 2.00 & 5.17 & 1.40 & M/H \\ \cmidrule(lr){1-7}
23 & Social med comp. & ban the use of AI on their platforms. & 1.91 & 5.00 & 1.61 & M/L \\ \cmidrule(lr){1-7}
24 & Government & fund community events to educate on media literacy. & 1.84 & 5.11 & 1.82 & M/L \\ \cmidrule(lr){1-7}
\multirow{2}{*}{25} & 
\multirow{2}{*}{Government} & 
ban AI companies from using public figures' & 
\multirow{2}{*}{1.80} & 
\multirow{2}{*}{4.60} & 
\multirow{2}{*}{2.17} & 
\multirow{2}{*}{M/L} \\ 
& & likeness in training data. & & & & \\\cmidrule(lr){1-7}
\bottomrule
\end{tabular}
\end{small}
\caption{Results from Survey 2 ranking stakeholder action pairs specific to harms from generative AI concerning overreliance on AI in regards to trustworthiness harms.}
\label{tab:Tr_OAI_results}
\end{table*}
\clearpage
\begin{table*}[t]
\begin{small}
\addtolength{\tabcolsep}{-0.08em}
\begin{tabular}{c|c|l|c|c|c|c}\toprule
 \multicolumn{7}{c}{\textbf{Well Being: Addiction} -- Stakeholder-Action Pair Ranking} \\
 \midrule
 \multirow{2}{*}{\textbf{}} &
 \multirow{2}{*}{\textbf{Stakeholder}} &
 \multicolumn{1}{c|}{\multirow{2}{*}{\textbf{Action}}} &
 \textbf{Priority} &
 \textbf{Agree} &
 \textbf{Consensus} &
 \textbf{Agree/} \\
  &
  &
  &
 (Mean  $\uparrow$) &
 (Mean $\uparrow$) &
 (Agree. SD $\downarrow$) &
 \textbf{Consen} \\
 \cmidrule(lr){1-7}
1 & News publishers & have strict quality control when using AI. & 2.60 & 6.20 & 0.68 & H/H \\ \cmidrule(lr){1-7}
\multirow{2}{*}{2} & 
\multirow{2}{*}{Schools} & 
emphasize human values and interactions   & 
\multirow{2}{*}{2.50} & 
\multirow{2}{*}{6.31} & 
\multirow{2}{*}{0.95} & 
\multirow{2}{*}{H/H} \\ 
& & over using AI. & & & & \\\cmidrule(lr){1-7}
3 & Government & require human oversight in news. & 2.5 & 6.00 & 1.71 & H/L \\ \cmidrule(lr){1-7}
\multirow{2}{*}{4} & 
\multirow{2}{*}{Tech companies} & 
include content warnings for potential   & 
\multirow{2}{*}{2.45} & 
\multirow{2}{*}{6.09} & 
\multirow{2}{*}{1.04} & 
\multirow{2}{*}{H/H} \\ 
& & negative mental health effects. & & & & \\\cmidrule(lr){1-7}
5 & Schools & limit access to generative AI tools. & 2.36 & 5.14 & 1.99 & M/L \\ \cmidrule(lr){1-7}
6 & Government & ensure accurate production of news. & 2.29 & 5.41 & 1.80 & M/L \\ \cmidrule(lr){1-7}
\multirow{2}{*}{7} & 
Public health & 
 raise awareness about mental health   & 
\multirow{2}{*}{2.29} & 
\multirow{2}{*}{5.86} & 
\multirow{2}{*}{1.61} & 
\multirow{2}{*}{M/L} \\ 
& officials & impacts of overuse of generative AI. & & & & \\\cmidrule(lr){1-7}
8 & News publishers & strengthen human element of news. & 2.27 & 5.91 & 1.38 & M/H \\ \cmidrule(lr){1-7}
9 & Ind. third parties & develop systems to stop misinformation. & 2.22 & 6.11 & 1.17 & H/H \\ \cmidrule(lr){1-7}
10 & Government & regulate AI companies. & 2.21 & 5.14 & 1.75 & M/L \\ \cmidrule(lr){1-7}
11 & Government & penalize people who abuse AI systems. & 2.19 & 5.44 & 1.79 & M/L \\ \cmidrule(lr){1-7}
12 & Schools & incorporate digital literacy programs. & 2.17 & 6.06 & 1.16 & H/H \\ \cmidrule(lr){1-7}
13 & Schools & limit students' access to AI. & 2.16 & 5.42 & 1.46 & M/H \\ \cmidrule(lr){1-7}
14 & Government & set standards to regulate content. & 2.14 & 4.86 & 2.35 & M/L \\ \cmidrule(lr){1-7}
\multirow{2}{*}{15} & 
Public health & 
develop specialized treatments for   & 
\multirow{2}{*}{2.07} & 
\multirow{2}{*}{5.27} & 
\multirow{2}{*}{2.31} & 
\multirow{2}{*}{M/L} \\ 
& officials & digital information addiction. & & & & \\\cmidrule(lr){1-7}
\multirow{2}{*}{16} & 
Independent & 
evaluate mental health effects of   & 
\multirow{2}{*}{2.06} & 
\multirow{2}{*}{5.33} & 
\multirow{2}{*}{1.75} & 
\multirow{2}{*}{M/L} \\ 
& third parties & consumption of generative AI outputs. & & & & \\\cmidrule(lr){1-7}
17 & Schools & promote healthy behavior using genAI. & 2.05 & 5.85 & 1.46 & M/H \\ \cmidrule(lr){1-7}
18 & News publishers & limit AI usage in news production. & 2.00 & 5.67 & 1.63 & M/L \\ \cmidrule(lr){1-7}
19 & Pub. health off. & programs to strengthen digital literacy. & 1.94 & 5.53 & 1.50 & M/L \\ \cmidrule(lr){1-7}
20 & News publishers & completely restrict AI generated stories. & 1.89 & 4.44 & 1.95 & M/L \\ \cmidrule(lr){1-7}
21 & Government & set standards to regulate ads in genAI. & 1.73 & 4.53 & 2.20 & M/L \\ \cmidrule(lr){1-7}
22 & Government & restrict AI generated news articles. & 1.69 & 4.31 & 1.89 & M/L \\ \cmidrule(lr){1-7}
23 & Tech companies & create healthy usage tools for AI tools. & 1.67 & 5.33 & 1.87 & M/L \\ \cmidrule(lr){1-7}
24 & Tech companies & audit competitors' algorithms. & 1.67 & 4.47 & 2.20 & M/L \\ \cmidrule(lr){1-7}
25 & Government & regulate harmful, AI generated content. & 1.67 & 4.08 & 2.19 & M/L \\ \cmidrule(lr){1-7}
26 & Tech companies & impose age limits on generative AI. & 1.56 & 3.31 & 1.82 & L/L \\ \cmidrule(lr){1-7}
27 & Tech companies & halt AI development until further research. & 1.53 & 3.53 & 2.56 & L/L \\ \cmidrule(lr){1-7}
28 & Tech companies & impose time limits for access to genAI tools. & 1.41 & 4.12 & 2.29 & M/L \\ \cmidrule(lr){1-7}
\bottomrule
\end{tabular}
\end{small}
\caption{Results from Survey 2 ranking stakeholder action pairs specific to harms from generative AI concerning addiction in regards to well being.}
\label{tab:WB_Ad_results}
\end{table*}
\clearpage


\subsection{Description of Impact Types}\label{sec:appendixA3}

The following descriptions of impact types are adapted from our previous work \cite{kieslich_anticipating_2024,barnett2024simulating}.

\textbf{Autonomy - Loss of Control.} This impact describes the loss of human autonomy over AI systems due to the ubiquity and inevitability of these systems. For example, scenarios describe how news consumers lose control over news consumption due to the increasing and subtle influence of generative AI applications in news.

\textbf{Labor - Unemployment.} This impact describes the impact on jobs that the implementation of AI in the news environment could have. Specifically, scenarios describe how people will lose their jobs as many of their tasks are replaced by generative AI. 

\textbf{Media Quality - Credibility/Authenticity.} This impact describes the loss of credibility that news outlets may experience as a result of implementing generative AI. Scenarios in this category describe the creation and distribution of fake news stories that result (after a public outcry) in a loss of credibility for the news outlet.

\textbf{Media Quality - Lack of Diversity/Bias.} This impact describes the loss of diversity in the news produced by generative AI. In particular, through personalization, the news consumer could be presented with homogeneous news that is tailored to the preferences and ideologies of the news reader. Scenarios in this category describe how news readers are sucked into their personalization algorithms, leading to silos in their worldviews.

\textbf{Media Quality - Lack of Fact Checking.} This impact refers to the consequences of a lack of fact-checking. Scenarios describe how unchecked false news can spread and have real-world consequences (e.g., for stock markets; civil unrest).

\textbf{Media Quality - Sensationalism.} This impact describes how generative AI can express emotional appeals that contribute to making news stories more sensational. Scenarios describe the real-world consequences for people affected by sensationalized stories, such as (untrue) allegations of corruption.

\textbf{Political - Fake News/Misinformation.} This impact describes how generative AI can facilitate the spread of fake news (e.g. due to sensationalism or hallucinations). Scenarios describe the blurring of fact and fiction due to the persuasive nature of AI-generated news stories. 

\textbf{Political - Manipulation.} This impact describes how generative AI can be used by malicious actors to manipulate public opinion. Scenarios describe how a story can be framed in different ways, creating friction between different ideological and political positions. 

\textbf{Trust - Overreliance on AI.} This category describes the negative effects that news consumers might experience as a result of overreliance on AI systems. For example, scenarios describe how news readers do not question and blindly rely on AI-generated fake news. This can have real-life consequences (e.g., investing in stocks).

\textbf{Well-Being - Addiction.} This impact describes how increasingly sophisticated AI algorithms have the potential to make users addicted. Scenarios describe how personalization systems have drawn people to their screens, resulting in a lack of human interaction.

\clearpage
\subsection{Example ``Policy Fact Sheet''}\label{sec:appendix_example_fact_sheet}

\begin{figure}[h]
    \centering
    \fbox{\includegraphics[width=0.85\linewidth]{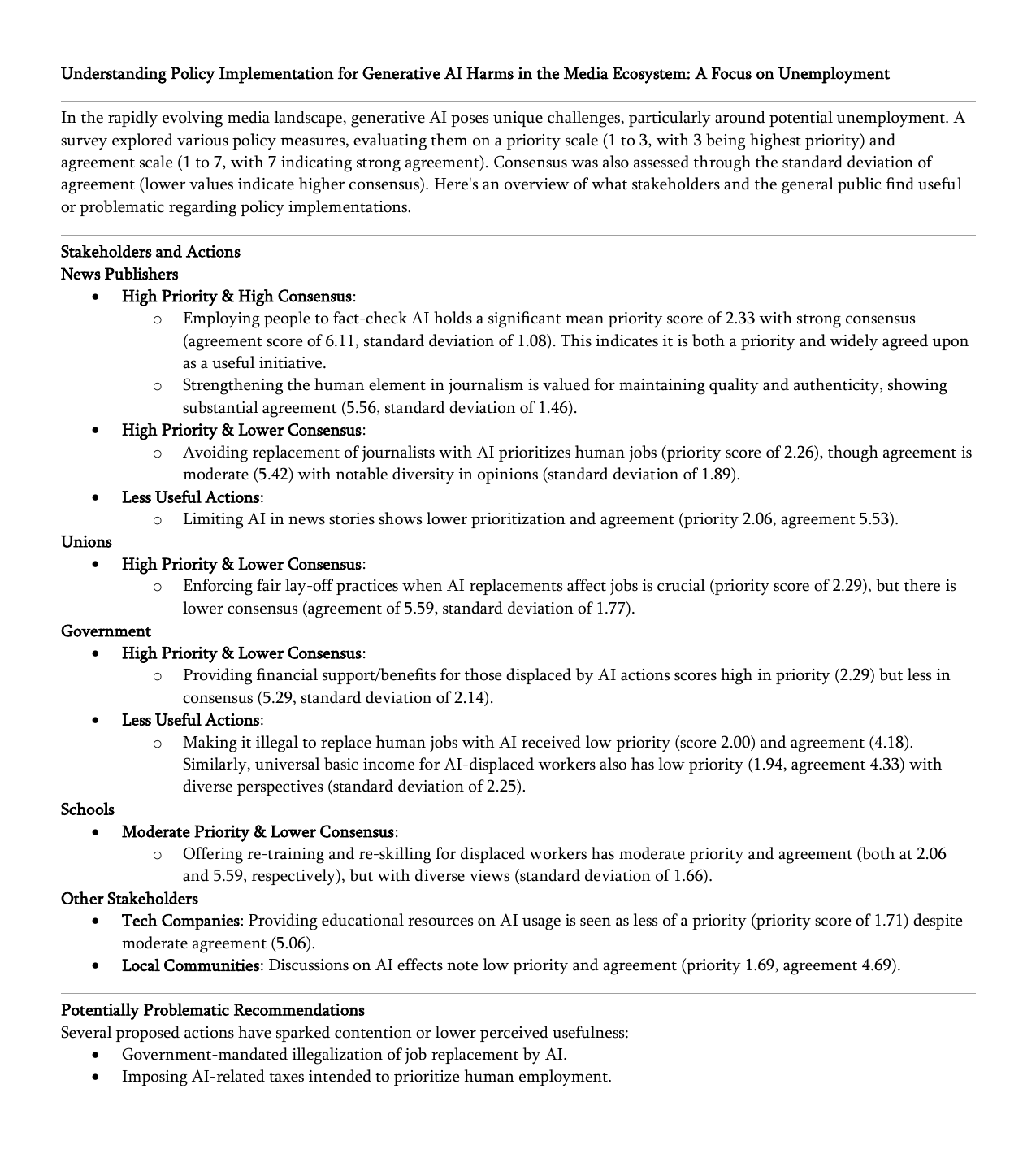}}
    \caption{Example ``Policy Fact Sheet'' for unemployment impacts generated using GPT-4o, prompt engineering, and the data from this study.}
    \label{fig:enter-label}
\end{figure}

\subsection{Prompting for Policy Fact Sheets}\label{sec:appendix_prompt_engineering}

The following prompt was used with gpt-4o using the OpenAI API.

``role'': ``system'', ``content'': ``You are a thoughtful assistant, skilled in synthesizing data into actionable insights for policy makers.''

``role'': ``user'', "content": 

Please create a short digestible document that would educate policy makers about what people find useful in terms of policy implementation for harms from generative AI in the media ecosystem related specifically to <impact type> in regards to <impact theme>. 

The following is a csv style table showing their mean priority score (higher is higher priority; scale 1-3), mean agreement (higher is stronger agreement; scale 1-7), standard deviation of agreement (lower is higher consensus), and categorical label of consensus (higher is better):"

<Data from the second survey for the specific impact type cleaned and ranked by mean priority, mean agreement score, standard deviation of agreement score, and categorical label for consensus. Presented in a comma separated format.>

Please create a one-page document highlighting these findings for policy-makers. Structure it by identifying what people find useful and what they don't find useful. 
Do not make policy recommendations.
Explain which brainstormed recommendations are both high priority and high consensus, highlight ones that are high priority but lower consensus, and identify elements people do not find useful. Structure this page around the stakeholders rather than the actions.
Identify stakeholders people want to take the action.
Make a separate section identifying recommendations people find potentially problematic. 
You can make insights from the consensus label, but never refer to them in the text.
A score greater than 6 indicates high agreement, especially if there is a higher consensus.
During the introductory context, mention that priority was scored out 3 with 3 being higher priority, agreement out of 7 with 7 being strongly agree, and standard deviation reported for agreement; describe how to interpret the scales.
High mean agreement scores mean high agreement. Low standard deviation scores for agreement mean high consensus.
When discussing diversity please mention the standard deviation score.

\end{document}